\documentclass[nofootinbib,preprint,onecolumn,superscriptaddress,amsmath,amssymb,floatfix,longbibliography]{revtex4-1}

\usepackage{soul}
\usepackage{amsmath}
\usepackage{graphicx}
\usepackage{dcolumn}
\usepackage{bbm}
\usepackage{bm}
\usepackage{subfigure}
\usepackage{amssymb}
\usepackage{xcolor}
\usepackage{subfigure}
\usepackage{makecell}
\usepackage{mathtools}
\usepackage{diagbox}
\usepackage[colorlinks,urlcolor=blue,linkcolor=blue,anchorcolor=blue,citecolor=blue]{hyperref}
\usepackage[bottom]{footmisc}
\usepackage{xparse}
\usepackage{tikz-feynman}
\usepackage{tikz-3dplot}

\usetikzlibrary{calc,snakes}

\tikzset{decoration={snake,amplitude=.4mm,segment length=1mm, post length=0mm,pre length=0mm}}

\usepackage{color}

\stepcounter{secnumdepth}
\stepcounter{tocdepth}

\newcommand{\bqa}{\begin{eqnarray}} 
\newcommand{\eqa}{\end{eqnarray}}
\newcommand{\nn}{\nonumber \\}
\newcommand{\eq}[1]{Eq. \eqref{#1}}

\begin{document}

\title{
Exact effective action for the O(N) vector model \\ in the large N limit
}
\author{
Han Ma}
\affiliation{\vspace*{0.1 cm} Perimeter Institute for Theoretical Physics, Waterloo, Ontario N2L 2Y5, Canada}
\author{Sung-Sik Lee}
\affiliation{\vspace*{0.1 cm} Perimeter Institute for Theoretical Physics, Waterloo, Ontario N2L 2Y5, Canada}
\affiliation{
Department of Physics $\&$ Astronomy, McMaster University,
1280 Main St. W., Hamilton, Ontario L8S 4M1, Canada }

\begin{abstract}

We present the Wilsonian effective action 
as a solution of the exact RG equation 
for the critical $O(N)$ vector model
in the large $N$ limit.
Below four dimensions,
the exact effective action 
can be expressed in a closed form as a transcendental function of two leading scaling operators with infinitely many derivatives. 
From the exact solution that describes the RG flow from a UV theory to the fixed point theory in the IR,
we obtain the mapping between UV operators and IR scaling operators. 
It is shown that IR scaling operators are given by sums of infinitely many UV operators with infinitely many derivatives.

\end{abstract}

\maketitle

\section{Introduction}

The renormalization group (RG) equation describes the change of effective theories as the energy cutoff is lowered\cite{Wilson1975renormalization,POLCHINSKI1984exactRG,WETTERICH1993evolutionequation}.
In particular, the infrared (IR) fixed points of the RG equations 
embody the notion of universality classes,  and play the central role in the classification of phases of matter  and our understanding of critical phenomena \cite{Kadanoff1966scaling,Callan1970Broken,Symanzik1970aa,Zomolodchikov1986irreversibility,Osborn1991Weyl,CARDY1988ctheorem,Komargodski2011aa,NAKAYAMA20151}.
In condensed matter physics, it is also important to understand the RG flow from microscopic theories usually defined on lattices to continuum theories at long distance scales.
However, understanding the exact RG flow of effective actions is generally hard for interacting theories.
The challenge lies in the fact that 
operators composed of arbitrary numbers of  fields and space derivatives are generated
under the exact RG flow.

If a theory comes with a small parameter such as $1/N$ with $N$ being the number of fields, a great deal of simplification arises.
The best understood example is the $O(N)$ vector model in the large $N$ limit.
In this theory, 
the exact RG equation for the 
 quantum effective potential 
 for spacetime-independent fields 
 can be written in a closed form thanks to the local potential approximation 
 that becomes exact in the large $N$ limit\cite{
WETTERICH1993evolutionequation,
morris,
1997PhLB..409..363D,
2006PhLB..632..571B}.
It allows for a systematic computation of the 
 effective potential in the power series of the field
\cite{morris,TETRADIS1994541,tetradis1996analytical,morris1997three,comellas1997on,berges2002non,moshe2003quantum,litim2011ising,metzner2012functional,knorr2020exact,ziebell2020existence,dupuis2021nonperturbative}.
However, the full scale dependent solution of the exact RG equation has not 
 been obtained yet.
At the IR fixed point of the exact RG equation, 
the formal expression of the quantum effective action 
has been written down as a functional of one collective variable being the O(N) singlet\cite{coleman1974spontaneous,moshe2003quantum}.
However, the full relation between the collective field and fundamental microscopic field in the O(N) vector representation is still unknown.

In this paper, we provide additional insights into the exact effective action of the $O(N)$ vector model.
The new results of our work are three-folded.
Firstly, we obtain the exact solution of the RG equation in the large $N$ limit in terms of the collective variable.
The IR limit of our scale dependent effective action is related to the effective action\cite{coleman1974spontaneous,moshe2003quantum} through the Legendre transformation.
Then, the saddle point equation allows us to express the collective field in terms of the fundamental fields.
Secondly, 
the full effective action is obtained not only at the IR fixed point but at general scales. 
This finds us the exact mapping between UV operators and IR scaling operators.
Here, UV operators refer to operators that can be added to a UV Lagrangian while
IR operators are the scaling operators that arise in the long-distance limit of the theory, meaning that they are the ones whose form do not change under the coarse graining and dilatation.
We note that the relation between UV operators 
and IR scaling operators 
 is highly non-trivial :
turning on a local operator in a lattice model generally amounts to turning on a linear superposition of many IR scaling operators in the continuum theory 
and conversely, an IR 
 scaling operator corresponds to a sum of multiple UV operators that involve infinitely many derivatives.
Based on the exact mapping derived from the solution of the RG equation,
we explicitly compute two leading IR scaling operators
 in terms of the UV fields.
As expected, the IR scaling operators involve arbitrarily many fundamental fields and derivatives.
Finally, we go beyond the large $N$ limit to compute the leading $1/N$ corrections to the effective action by including fluctuations of the collective field.

To achieve these goals,
we use a recently developed  quantum RG scheme \cite{lee2012background,lee2014quantum,lee2016horizon}, which allows us to keep track of the RG flow starting from a UV theory
to the O(N) Wilson Fisher fixed point\footnote{
The Wilsonian effective action that satisfies the  exact Polchinski RG equation\cite{POLCHINSKI1984exactRG} is related to the 1PI effective action through the Legendre transformation.}.
As an exact reformulation of the Wilsonian RG, 
the quantum RG significantly reduces the number of operators that need to be included along the exact RG flow as couplings are promoted to dynamical variables.
The information of the operators that are not explicitly kept within the RG flow is encoded in  fluctuations of the dynamical couplings. 
We note that most of the results obtained in our paper 
for the $O(N)$ vector model
through the quantum RG 
can be in principle obtained through alternative methods that use collective fields.
However, the quantum RG can be readily extended to more complicated theories such as matrix models\cite{lee2012background,ma2020constraints} for which the method of the collective field used for the vector model is not applicable.

The outline of this paper is as follows. In Sec.~\ref{sec:QRG}, we apply the quantum RG method to the O(N) vector model. Subsequently, Sec.~\ref{sec:IR_action} gives us the effective action in the IR. In the large N limit, we are able to express this action in terms of the UV field and more surprisingly it can be written in a closed form in terms of scaling operators as discussed in Sec.~\ref{sec:scaling_op}. Furthermore, in Sec.~\ref{sec:correlation_function}, we obtain the generating function from the effective action which allows us to compute correlation functions. Sec.~\ref{sec:conclusion} summarizes our main results.

\section{Quantum Renormalization group \label{sec:QRG}}

Consider the partition function of the $O(N)$ vector model 
defined on a $D$-dimensional Euclidean lattice,
\begin{eqnarray}
Z=\int \mathcal{D} \phi  e^{-\frac{m^2}{2}\left[ \sum_{i} \phi_{i}^2
+\sum_{ij} M_{ij} \phi_{i}\phi_{j}
\right]- \frac{\lambda}{N}\sum_{i }  ( \phi_{i}^2)^2},
\label{eq:Z_UV}
\end{eqnarray}
where $\phi^a_i$ is the fundamental field  with flavour $a=1,2,..,N$ defined at site $i$, 
$\phi_i \phi_j \equiv \sum_a \phi^a_i \phi^a_j$,
$m$ is the on-site mass,
$-\frac{m^2}{2}M_{ij}$ is the hopping amplitude between site $i$ and $j$, and
$\lambda$ is the on-site quartic interaction.
In describing the RG flow,
it is convenient to divide the action 
into a reference action and 
a deformation added to the reference action.
Here we choose 
the insulating fixed point action,
$S_{ref}=\frac{m^2}{2} \sum_i\phi_i^2$
as the reference action. 
The hopping (kinetic) term and  the quartic interaction
are regarded as deformations, 
$S_{1} = \frac{\lambda}{N}\sum_{i }  ( \phi_{i}^2)^2
+\frac{m^2}{2}\sum_{ij} M_{ij} \phi_{i}\phi_{j}
$, which is not necessarily small.
Depending on its magnitude, 
the theory can flow to 
one of the fixed points
in the long-distance limit.
Above two dimensions,
the possible fixed points include
the insulator, 
the critical point,
and the long-range ordered state. 
Choosing a reference theory is equivalent to picking a point in the space of theories as the origin.
Physics does not depend on this choice\footnote{
In the appendix, we present the effective action obtained with the choice of the critical Gaussian fixed point theory as the  reference action.
}.

In quantum RG, 
every local action is associated with 
a short-ranged entangled quantum state\cite{lee2016horizon}.
For the reference action and the deformation, we introduce
$|S_{ref} \rangle = \int \mathcal{D}\phi ~ e^{-
S_{ref}
} |\phi\rangle$
and
$|S_{1} \rangle = 
\int \mathcal{D}\phi ~ e^{-
S_{1} }
|\phi\rangle $, respectively, 
where
$| \phi \rangle $ is the basis state defined in the $D$ dimensional lattice with the inner product
$\langle \phi' | \phi \rangle = \prod_{i,a} \delta( \phi^{'a}_i - \phi^a_i )$.
These are $D$-dimensional states whose wavefunctions are given by the exponentials of the actions.
Then the partition function is an overlap between the two states, 
\begin{eqnarray}
Z=\langle S_{ref}  |S_{1} \rangle.
\label{eq:Z_quantum_states}
\end{eqnarray}
The renormalization group transformation can be  described as a quantum evolution  of states associated with actions.
An infinitesimal RG transformation  is implemented by a quantum evolution operator $e^{-dz \hat H}$, 
where $\hat H$ is an RG Hamiltonian and $dz$ is the infinitesimal RG time. 
As will be shown later, $z$
corresponds to the logarithmic length scale. 
Since we have chosen the ultra-local action as the reference theory, 
it is natural to use 
a coarse graining transformation 
under which the insulating fixed point action 
is invariant.
The RG Hamiltonian that leaves the 
insulating fixed point invariant 
is given by
\begin{eqnarray}
\hat{H} =\sum_i\left[ 
\frac{1}{m^2}\hat{\pi}_i^2 +
i \hat{\phi}_i \hat{\pi}_i  
\right].
\label{eq:RGH}
\end{eqnarray}
$\hat \pi_i^a$ is the conjugate momentum of $\hat \phi_i^a$ that satisfies 
the commutation relation,
$[\hat{\phi}_i^a,\hat{\pi}_j^b]=i\delta_{ij} \delta_{ab}$. More details about its derivation can be found in Appendix \ref{app:RG_Hamiltonian_O(N)}.
This RG Hamiltonian is the generator of the exact Polchinski equation\cite{POLCHINSKI1984exactRG} in real space\cite{lee2016horizon}.
$e^{- dz \hat{\pi}_i^2/m^2 }$
has the effect of partially integrating out modes at each site 
without reducing the number of sites.
The remaining `low-energy' degrees of freedom have less fluctuations and hence a larger mass $m e^{dz}$.
On the other hand, 
$e^{ -i dz \hat{\phi}_i \hat{\pi}_i  } $
scales the field as 
$\phi_i \rightarrow e^{-dz} \phi_i$ 
so that the increased mass in the reference action is put back to the original value.
Because $\hat{H}^\dag |S_{ref}\rangle =0$,
the partition function is invariant under the insertion of the RG evolution operator between the overlap,
\begin{eqnarray}
Z = \langle S_{ref}  |e^{-\hat{H} z^\ast}|S_{1} \rangle,
\label{eq:SHzS}
\end{eqnarray}
where $z^\ast$ increases along the RG flow.
Applying the RG evolution operator to 
$| S_{1} \rangle$ in \eq{eq:SHzS},
one obtains the state at scale $z^\ast$,
$| S_1^{z^*} \rangle =
e^{- \hat{H} z^\ast} 
|S_{1}\rangle $.
The state gives the renormalized deformation at scale $z^*$ as
$S_1^{z^*}= -\ln \langle \phi | S_1^{z^*} \rangle$
from the state-action correspondence\cite{lee2016horizon}.
$S_1^{z^\ast}$ corresponds to the effective action obtained with the IR cutoff
$m^2 \frac{ e^{-2z^\ast}}{1-e^{-2z^\ast}}$ (Appendix \ref{app:EffectiveAction}).
The renormalized action includes 
infinitely many new higher order interactions,
$
\sum_{m=2}^\infty
J_{i_1,j_1;..;i_m,j_m}(z^\ast) 
\frac{1}{N^{m-1}}
\prod_{k=1}^m (\phi_{i_k} \phi_{j_k})
$,
which remain important
in the large $N$ limit.
In quantum RG, 
one does not need to keep track of all operators.
This simplification arises from the fact that
i) the space of theories is viewed as a Hilbert space,
and
ii) the Hilbert space of $O(N)$ invariant states 
can be spanned by basis states labeled by  the hopping field only,
\bqa
| t \rangle =\int \mathcal{D}\phi~ e^{
 i\sum_{ij}t_{ij}\phi_i\phi_j
}|\phi\rangle.
\label{eq:t}
\eqa
Therefore,
$| S_1^z \rangle$ can be expressed as a linear superposition of 
$| t \rangle $ at all $z$,
and one only needs
to keep track of the hopping fields
along the RG flow.
The price to pay is to sum over 
all RG paths for the hopping fields.
In other words,
the exact Wilsonian RG flow
defined in the space of an infinite tower
of higher order couplings
is expressed as a path integration 
over the dynamical hopping field only.
The $\beta$-functions are then replaced with an action 
that governs the dynamics 
of the $z$-dependent hopping field defined 
in a  $(D+1)$-dimensional bulk
\cite{LEE2012holographicmatter,lee2012background,lee2014quantum},
where the extra direction corresponds to the length scale $\ln z$.
The fact that the $\beta$-functions 
for all higher order couplings can be encoded in the dynamics of 
a much smaller subset of couplings 
also implies that 
the original $\beta$-functions 
are highly constrained
even 
in the presence of  arbitrary irrelevant deformations 
\cite{ma2020constraints}. 
In the phase space path integral representation,
$| S_1^{z^*} \rangle$ can be written as
\begin{eqnarray}
| S_1^{z^*} \rangle 
&=&
\int \mathfrak{D} t~\mathfrak{D}p ~
e^{
-N S_{UV}
-N S_{bulk}
} 
|
t^{z^\ast}
\rangle.
\label{eq:S1z}
\end{eqnarray}
Here, 
$\mathfrak{D}t  = \prod_z \mathcal{D}t^z$ and $\mathfrak{D}p = \prod_z \mathcal{D}p^z$
represents the sum over all RG paths.
$t_{ij}^z$ is the $z$-dependent
dynamical hopping fields between site $i$ and $j$.
$p_{ij}^z$ is the conjugate variable
that corresponds to the operator 
$\frac{1}{N}(\phi_i \phi_j)$. As derived in Appendix \ref{app:bulk_theory},
$S_{bulk}$ is the action
that determines the weight of 
each RG path,
\begin{eqnarray}
&& S_{bulk} = 
\int_0^{z^\ast}    dz
\Biggl[
i\sum_{ij}p^z_{ij}\partial_z t^z_{ij}-
\frac{2i}{m^2}\sum_k  {t}^z_{kk}  
 +2i \sum_{kl} {t}^z_{kl} {p}^{z}_{kl}
+\frac{4}{m^2} \sum_{kji}
{t}^z_{ik}
{t}^z_{kj} 
{p}^{z}_{ij}
\Biggr].
\label{eq:Sbulk}
\end{eqnarray}
$S_{UV}
=\sum_{ij}(it^0_{ij}+\frac{m^2}{2}M_{ij}) p^0_{ij}+\lambda\sum_{i }   (p_{ii}^0)^2$ is the action for the fields defined at $z=0$. 
It imposes a dynamical boundary condition for the interacting $O(N)$ model\cite{Witten:2001ua,KLEBANOV2002213}.
The quadratic term in $p^0_{ii}$ allows $t^0_{ii}$ to have non-trivial fluctuations at the UV boundary.
On the contrary, $S_{bulk}$ is linear in $p_{ij}^z$\cite{ dolan},
and $t_{ij}^z$ in the bulk is fixed by $t_{ij}^0$.
This fact 
is a special property of 
vector models 
in which the complete basis states 
are Gaussian in the fundamental field
as is shown in \eq{eq:t}
\cite{Das:2003vw,PhysRevD.83.071701,Leigh:2014tza}. 
For matrix models, 
basis states are non-Gaussian, 
and there exist non-trivial fluctuations 
of dynamical couplings 
both at the UV boundary and in the bulk\cite{lee2012background}.
$S_{bulk}$ in  \eq{eq:Sbulk} is related to the one derived in Ref. \cite{lee2016horizon} through a similarity transformation. 
The bulk theory is finite and well-defined as the original  $D$-dimensional field theory is  properly regularized\cite{LEE2012holographicmatter,lee2016horizon}. 

\section{IR deformation in the large N limit \label{sec:IR_action}}

The bulk path integration 
in \eq{eq:S1z}
can be readily performed. 
This allows one to write 
$| S_1^{z^*} \rangle $
in terms of the integration over 
$t^0$ and $p^0$ only,
\begin{eqnarray}
| S_1^{z^*} \rangle 
&=&
\int \mathcal{D} t^0 \mathcal{D}p^0
e^{-N \left[S_{UV}
-
\sum_i 
\int_0^{z^\ast} dz 
\frac{2i { t}^z_{ii}}{m^2}\right] }
|
t^{z^\ast}
\rangle.
\label{eq:S1zast0}
\end{eqnarray}
Here  $  t_{ij}^z$ is the solution  of 
$ \partial_z t_{ij}+2t_{ij}-\frac{4 i }{m^2}\sum_k t_{ik}t_{kj}  = 0 $
given by
$i  {\bf  t}^{z} 
= (i{\bf  t}^{0})  \left[   e^{2z} -\frac{2}{m^2} 
(e^{2z}-1)(i{\bf  t}^{0})\right]^{-1}$,
where 
${\bf t}^z$ is a square matrix
whose elements are $\{ t_{ij}^z \}$
and
${\bf t}^0$ is the hopping matrix at $z=0$.
The fluctuations of the hopping fields
encode higher order operators 
in \eq{eq:S1zast0}.
At $z^\ast = 0$,
the action for $t^0$
consists of $S_{UV}$ only.
Because $S_{UV}$ is Gaussian,
integrating over $t^0$ and $p^0$ 
reproduces the quartic interaction
at $UV$.
At $z^\ast > 0$, however, 
the action for $t^0$
becomes non-Gaussian
because of the 
the on-shell bulk 
action in \eq{eq:S1zast0}.
The non-Gaussian fluctuation 
of the hopping fields 
is what captures higher order couplings
in $z^\ast > 0$.

With this picture in mind,
we proceed to compute the fixed point action 
that the theory flows into 
in the large $z^*$ limit.
In the large $N$ limit,
the remaining integration
can be replaced with the $\phi$-dependent
saddle-point. 
In \eq{eq:S1zast0},  
$S_{UV}$  is linear  in $p^0_{ij}$ 
for  $i \neq j$,
and the off-diagonal elements of 
$t_{ij}^0$ is
fixed by the hopping field at UV :
$-it^0_{ij} = \frac{m^2}{2} M_{ij}$ for  $i \neq j$.
This leaves only the diagonal elements of 
$t^0_{ii}$ and $p^0_{ii}$ 
non-trivial.
To isolate the deviation of 
$-it^0_{ij}$ from
$ \frac{m^2}{2} {\bf M}$ for  $i = j$,
we write
$-i{\bf t}^0 
=   \frac{m^2}{2} ( {\bf M} +  {\bf X})$,
where ${\bf X}_{ij} = X_i \delta_{ij}$ is a diagonal matrix.
The deformation at $z^\ast$,
$S_{1}^{z^\ast}=  -  
\log \langle \phi | S_1^{z^*} \rangle$
becomes 
\begin{eqnarray}
&&  S_{1}^{z^\ast} 
= 
\frac{m^2}{2} \frac{1}{e^{2z^\ast}-1}
\sum_{ij} \left[  I- \frac{e^{2z^\ast}}{(e^{2z^\ast}-1)}\frac{1}{{\bf K}+{\bf X}}
\right]_{ij} \phi_{i}\phi_{j}  \nn
&&
+ \frac{N}{2} \mbox{tr} \log 
\left[
(e^{2z^\ast}-1)
( {\bf K}+{\bf X})
\right]
-\frac{N m^4}{16 \lambda}\sum_{i } X_{i}^2
\label{eq:Stot}
\end{eqnarray}
up to a constant,
where  
${\bf K}=
\left[   {\bf M}+\frac{e^{2z^\ast}}{(e^{2z^\ast }-1)} I \right]$
 with $I$ being the identity matrix, and ${\bf X}$ is the saddle point solution satisfying
\begin{eqnarray}
&&m^2 e^{2z^\ast }
\sum_{kl} 
({\bf K}+ {\bf X})^{-1}_{ik}
(\phi_k \phi_l)
({\bf K}+ {\bf X})^{-1}_{li}
+
N(e^{2z^\ast }-1)^2 \left[
({\bf K}+ {\bf X})^{-1}_{ii}- \frac{m^4 }{4 \lambda}   {X}_{i}  \right]
=0 
\label{eq:eqforX0}
\end{eqnarray}
for each and every site $i$ (not summed over).
The solution of \eq{eq:eqforX0} can be written in powers of 
$( \phi_i \phi_j)$   as
${X}_{i}(z) = 
\sum_{m=0}^\infty
\sum_{j_1,..,j_m} \sum_{k_1,..,k_m}
{x}^{j_1-i, j_2-i,..,j_m-i}_{k_1-i, k_2-i, ..,k_m-i}(z)$
$( \phi_{j_1} \phi_{k_1})
$$( \phi_{j_2} \phi_{k_2})
..
$$( \phi_{j_m} \phi_{k_m})$,
where 
the rank $2m$ tensor 
${x}^{j_1-i, j_2-i,..,j_m-i}_{k_1-i, k_2-i, ..,k_m-i}(z)$
does not depend on $i$ separately 
because of the translational invariance.
The zeroth order coefficient ${x}(z)$,
which is a $z$-dependent function,
satisfies the self-consistent equation,
\begin{eqnarray}
\label{eq:equation_x}
&&
\left[
\frac{1}{
{x}(z)I + {\bf K}
} \right]_{ii}
- \frac{m^4}{4 \lambda}  {x}(z) = 0.\\ \nonumber
\end{eqnarray}
For $D>2$, 
${x}(z) = 
x_0
+  x_2 e^{-2z^\ast} 
+ \mathcal{O}(e^{-4z^\ast})$, 
where $x_0 \propto \lambda \Lambda^{D-2}$, and
$\Lambda$ is the large momentum cutoff  at $z^\ast=0$ (Appendix \ref{app:solution_x}).

In the continuum, 
${\bf M}_{ij}$ can be written as
$M(r_i,r_j) =  \left[ a -  
\frac{\nabla_j^2}{m^2} + O(\frac{\nabla_j^4}{m^4}) \right]  \delta(r_i-r_j) $, 
where $a$ is a constant fixed by the UV hopping, and
the coefficient of $\nabla^2$ is set to be $-\frac{1}{m^2}$ without loss of generality.
The bare mass of the UV theory is given by $m^2(1 + a)$,
and $a$ can be used to tune the system 
across the insulator 
to symmetry breaking phase transition.
The first term in 
\eq{eq:Stot} can be written as
$\frac{m^2}{2} \int dr \phi(r) L \phi(r)$,
where 
$L$ is a differential operator 
which 
in the large $z^\ast$ limit 
takes the following form 
to the leading order in $\phi$ and the number of derivatives : 
$L= \frac{ 
e^{2z^\ast } \Big( 1+a+{x}(z^\ast) -  \frac{\nabla^2}{m^2} + O( \frac{\nabla^4}{m^4}) + O( \phi^2 )  \Big)
}{
(e^{2z^\ast }-1) \Big(  1+a+{x}(z^\ast)  -  \frac{\nabla^2}{m^2}
+ O( \frac{\nabla^4}{m^4}) + O( \phi^2 ) \Big) + 1 }$. Next, we would like to discuss different cases characterized by zero or nonzero $\delta$.

\subsection{Massive theory \label{sec:massive}}

For $\delta \equiv 1+a+{x}_0 > 0$, 
$L = 1 + O(e^{-2z})$ 
in the large $z^*$ limit.
In this case, the first term in \eq{eq:eqforX0}
is suppressed by $e^{-2 z^\ast}$ 
compared to the second term 
at large $z^\ast$.
Consequently,
${X}_{i}$ becomes independent of $\phi$,
and its saddle-point equation reduces to  \eq{eq:equation_x}.
In this case, the effective action is simply given by $S_{ref}$.
This shows that small hopping and interaction are irrelevant at the ultra-local insulating fixed point. 
With the strength of hopping increased, 
the critical point 
and the long-range ordered state  can be reached \cite{lee2016horizon}.

\subsection{Critical theory \label{sec:critical}}
At the critical point,
the fixed point action becomes qualitatively different.
With $\delta = 0$,
$\Sigma_{z^\ast} \equiv e^{2z^\ast} \Bigl( {x}(z^\ast) +a+1 \Bigr)+1$ approaches $\tilde \Sigma\equiv x_2 + 1$ in the large $z^\ast$ limit.
In this case,
$L$ is local only at length scale larger
than $\frac{e^{z^\ast}}{ \Sigma_{z^\ast}^{1/2} m}$.
This implies that
at the critical point
the range of  the renormalized hopping keeps increasing 
without a bound with increasing $z$\cite{lee2016horizon}.
In order to have a well-defined large $z^\ast$  limit at the critical point,
one has to scale the coordinate and the field as
$\tilde{r}=re^{-z^\ast}$,
$\tilde{\phi}_{\tilde{r}}=
\phi_r e^{\frac{D}{2}z^\ast}$.
This shows that the RG parameter $z$  indeed plays the role of the logarithmic length  scale. Accordingly, the rescaling of the effective action and the saddle point equation is carried out in Appendix \ref{app:rescaling}.
At the critical point,
the effective action in 
\eq{eq:Stot} can be written 
in terms of the scaled variables as
\begin{widetext}
\begin{eqnarray}
\label{eq:action_rescaling}
&& \tilde{S}_{1}^{z^\ast} 
=
-\frac{1}{2}m^2 
\int d^D \tilde r d^D \tilde r' 
\left[\Big( \tilde{\bf T}+{\bf X}'\Big)^{-1}_{\tilde{r}\tilde{r}'}\tilde{\phi}_{\tilde{r}} \tilde{\phi}_{\tilde{r}'}
\right]
+\frac{N}{2}
\int d^D \tilde r 
\left[
\log \left(\tilde{\bf T}+{\bf X}' \right) \right]_{\tilde{r}\tilde{r}}  \\
&& ~~~
-  N \int d^D \tilde r  ~ 
\Big\{ 
\frac{m^4 }{16 \lambda} e^{(D-4)z^\ast}({X}'_{\tilde{r}}-x_2)^2
+ 
\frac{ 1}{2} 
\left[\tilde{\bf T}+x_2I\right]^{-1}_{\tilde{r}\tilde{r}}
({X}'_{\tilde{r} }-x_2)+ \frac{\lambda }{m^4} e^{(4-D)z^\ast}\Big(\left[\tilde{\bf T}+x_2I\right]^{-1}_{\tilde{r}\tilde{r}}\Big)^2
\Big\}.\nonumber
\end{eqnarray}
\end{widetext}
Here,
$\tilde{\bf T}_{\tilde{r}\tilde{r}'}=
\int^{\Lambda e^{z^\ast}} \frac{d^D \tilde{Q}}{(2\pi)^D}e^{i\tilde{Q}(\tilde{r}-\tilde{r}')}\left[ 
\frac{\tilde{Q}^2}{m^2} + 1 \right] $.
${ \bf X}'_{\tilde r \tilde r'} =  
\delta( \tilde r - \tilde r')
{X}'_{\tilde r}$,
where 
${X}'_{\tilde r } =  
e^{2z^\ast} 
\left(  {X}_{r} - {x}_0  \right)$.
Modulo constant term, the effective action is finite in the large $z^\ast$ limit\footnote{
The second to last term in \eq{eq:action_rescaling}
is UV divergent in the large $z^\ast$ limit because 
$\tilde{\bf T}_{\tilde{r}\tilde{r}}$
evaluates the matrix element
at a coincident point.
However, this UV divergence 
is cancelled by the  
${ \bf X}'$-linear term in
$\log \left(\tilde{\bf T}+{\bf X}' \right)$,
and the fixed point action is UV finite. }.
The equation for 
${ \bf X}'$ becomes
\begin{eqnarray}
\label{eq:saddle_point_rescaling}
&&
\frac{m^2}{N}
\int d^D \tilde{r}_1 d^D \tilde{r}_2
\left[\tilde{\bf T}+{\bf X}'\right]^{-1}_{\tilde{r}\tilde{r}_1}
\tilde{\phi}_{\tilde{r}_1} \tilde{\phi}_{\tilde{r}_2} 
\left[\tilde{\bf T}+{\bf X}'\right]^{-1}_{\tilde{r_2}\tilde{r}}
\\
&& + \left[\tilde{\bf T}+{\bf X}'\right]^{-1}_{\tilde{r}\tilde{r}}
-\left[
\tilde{\bf T} 
+  x_2 I 
\right]^{-1}
_{\tilde{r}\tilde{r}} 
= \frac{m^4 
e^{(D-4)z^\ast}
}{4 \lambda}  
( {X}'_{\tilde{r}}
-   x_2
).  \nonumber
\end{eqnarray}
The $z^\ast$ dependence of its solution in the large $z^\ast$ limit is determined by the sign of $D-4$.

\subsubsection{$D>4$}
For $D>4$, the last term in Eq.~(\ref{eq:saddle_point_rescaling}) grows exponentially while other terms remain order one. 
This forces 
${X}'_{\tilde{r}}= x_2$ in the large $z^\ast$ limit. 
In this case, the deformation reduces
to a simpler form,
\begin{eqnarray}
\tilde{S}_{1}^{z^\ast} 
&=&-\frac{m^2}{2} \int d^D \tilde{r} \tilde{\phi}_{\tilde{r}} \frac{1}{-\tilde{\nabla}^2/m^2+\tilde \Sigma }\tilde{\phi}_{\tilde{r}} + \mathcal{O}(e^{-(D-4)z^\ast}\tilde{\phi}^4).
\label{eq:action_gaussian}
\end{eqnarray}
As $z^\ast$ increases, the quartic interaction of $\tilde{\phi}$ decays exponentially,
and the action approaches the Gaussian form in $\tilde \phi$.\footnote{
At finite $z^\ast$, the small but non-zero quartic term generates 
a mass renormalization.
This is why the apparent mass term 
$(\tilde \Sigma -1)$ does not vanish in general at the  critical point. 
If one starts with the Gaussian theory at UV, $(\tilde \Sigma -1) = 0$.
}

\subsubsection{$D<4$}
For $D<4$,
the last term in Eq.~(\ref{eq:saddle_point_rescaling}) can be dropped, and 
${X}'_{\tilde r } $
depends on $\tilde \phi$.
It can be
computed order by order in 
$(\tilde{\phi}_{\tilde r} \tilde{\phi}_{\tilde r'})$ as
${X}_{\tilde r}'=x_2+ \sum_{k=1}^\infty 
{X}^{(k)}_{\tilde r}$,
where
${X}^{(k)}_{\tilde r} \propto 
(\tilde{\phi} \tilde{\phi})^k$. We can define the Feynman rules as
\begin{eqnarray}
\frac{m}{\sqrt{N}}\tilde{\phi}_{\tilde{r}}^a &=&
\begin{tikzpicture}[baseline={([yshift=-4pt]current bounding box.center)}]
\node at (0pt,0pt) [circle,fill,inner sep=1pt]{};
\end{tikzpicture}~,
\nonumber\\
(\tilde{\bf T}+x_2I)^{-1}_{\tilde{r}\tilde{r}'} &=& \begin{tikzpicture}[baseline={([yshift=-4pt]current bounding box.center)}]
\coordinate (v1) at (-12pt, 0pt);
\coordinate (v2) at (12pt,0pt);
\draw[thick](v1)--(v2);
\node at (-18pt,0pt) {\scriptsize $\tilde{r}$};
\node at (18pt,0pt) {\scriptsize $\tilde{r}'$};
\end{tikzpicture} ~,
\nonumber\\
%  &=&
\left[(\tilde{\bf T}+x_2I)^{-1}\right]^2_{\tilde{r}\tilde{r}'} &=&
\begin{tikzpicture}[baseline={([yshift=-4pt]current bounding box.center)}]
\coordinate (v1) at (-12pt, 0pt);
\coordinate (v2) at (12pt,0pt);
\draw[thick,decorate](v1)--(v2);
\node at (-18pt,0pt) {\scriptsize $\tilde{r}$};
\node at (18pt,0pt) {\scriptsize $\tilde{r}'$};
\end{tikzpicture}~.
\end{eqnarray}
The straight line represents the propagator of the free boson field. In the momentum space, it is $G_\phi (p)= p^{-2}$. While the wavy line stands for the propagator of the singlet field which is given by $G_{X} (p) \propto p^{4-D}$. 
Diagrammatically, the first and second order terms contributing $X'_{\tilde{r}}$ are given by
\begin{eqnarray}
{X}_{\tilde r}^{(1)}
&=& 
\begin{tikzpicture}[baseline={([yshift=-4pt]current bounding box.center)}]
\coordinate (v1) at (-12pt, 6pt);
\coordinate (v2) at (-12pt,-6pt);
\coordinate (v3) at (0pt,0pt);
\coordinate (v4) at (12pt,0pt);
\draw[thick,decorate](v3)--(v4);
\draw[thick](v1)--(v3);
\draw[thick](v2)--(v3);
% \node at (-18pt,0pt) {\scriptsize $r$};
\node at (18pt,0pt) {\scriptsize $\tilde{r}$};
\node at (v1) [circle,fill,inner sep=1pt]{};
\node at (v2) [circle,fill,inner sep=1pt]{};
\end{tikzpicture}
~,
\nonumber\\
{X}_{\tilde r}^{(2)}
&=& -2~
\begin{tikzpicture}[baseline={([yshift=-4pt]current bounding box.center)}]
\coordinate (v1) at (-12pt, 6pt);
\coordinate (v2) at (-12pt,-6pt);
\coordinate (v3) at (0pt,0pt);
\coordinate (v4) at (12pt,0pt);
\coordinate (v5) at (24pt,-6pt);
\coordinate (v6) at (24pt,6pt);
\coordinate (v7) at (12pt,12pt);
\coordinate (v8) at (36pt,6pt);
\draw[thick,decorate](v3)--(v4);
\draw[thick](v1)--(v3);
\draw[thick](v2)--(v3);
\draw[thick](v4)--(v5);
\draw[thick](v4)--(v6);
\draw[thick](v6)--(v7);
\draw[thick,decorate](v6)--(v8);
% \node at (-18pt,0pt) {\scriptsize $r$};
\node at (40pt,6pt) {\scriptsize $\tilde{r}$};
\node at (v1) [circle,fill,inner sep=1pt]{};
\node at (v2) [circle,fill,inner sep=1pt]{};
\node at (v5) [circle,fill,inner sep=1pt]{};
\node at (v7) [circle,fill,inner sep=1pt]{};
\end{tikzpicture}
~+~
\begin{tikzpicture}[baseline={([yshift=-4pt]current bounding box.center)}]
\coordinate (v1) at (-12pt, 6pt);
\coordinate (v2) at (-12pt,-6pt);
\coordinate (v3) at (0pt,0pt);
\coordinate (v4) at (12pt,0pt);
\coordinate (v5) at (24pt,0pt);
\coordinate (v6) at (18pt,10pt);
\coordinate (v7) at (12pt,12pt);
\coordinate (v8) at (36pt,0pt);
\coordinate (v9) at (48pt,6pt);
\coordinate (v10) at (48pt,-6pt);
\coordinate (v11) at (18pt,20pt);
\draw[thick](v1)--(v3);
\draw[thick](v2)--(v3);
\draw[thick,decorate](v3)--(v4);
\draw[thick](v4)--(v5);
\draw[thick](v4)--(v6);
\draw[thick](v5)--(v6);
\draw[thick,decorate](v5)--(v8);
\draw[thick](v8)--(v9);
\draw[thick](v8)--(v10);
\draw[thick,decorate](v6)--(v11);
% \node at (-18pt,0pt) {\scriptsize $r$};
\node at (18pt,25pt) {\scriptsize $\tilde{r}$};
\node at (v1) [circle,fill,inner sep=1pt]{};
\node at (v2) [circle,fill,inner sep=1pt]{};
\node at (v9) [circle,fill,inner sep=1pt]{};
\node at (v10) [circle,fill,inner sep=1pt]{};
\end{tikzpicture}~.
\label{eq:X^1}
\end{eqnarray}
More details are given in Appendix \ref{app:solution_X_phi}.
${X}^{(k)}$ is well defined for all $k$
in terms of the scaled variables in the large $z^\ast$ limit.

We can further expand \eq{eq:action_rescaling} in terms of $\tilde{\phi}_{\tilde{r}}$ field. Up to the quartic order, we can get
\begin{eqnarray}
\frac{1}{N}\tilde{S}^{z^\ast}_1 &=& -\frac{1}{2} \int d^D \tilde{r} d^D \tilde{r}'(
\begin{tikzpicture}[baseline={([yshift=-4pt]current bounding box.center)}]
\coordinate (v1) at (-12pt, 0pt);
\coordinate (v2) at (12pt,0pt);
\draw[thick](v1)--(v2);
\node at (-18pt,0pt) {\scriptsize $\tilde{r}$};
\node at (18pt,0pt) {\scriptsize $\tilde{r}'$};
\node at (v1) [circle,fill,inner sep=1pt]{};
\node at (v2) [circle,fill,inner sep=1pt]{};
\end{tikzpicture})+\frac{1}{4} \int d^D \tilde{r}_1 d^D \tilde{r}_2d^D \tilde{r}_3 d^D \tilde{r}_4 (\begin{tikzpicture}[baseline={([yshift=-4pt]current bounding box.center)}]
\coordinate (v1) at (-12pt, 0pt);
\coordinate (v2) at (12pt,0pt);
\coordinate (v3) at (0pt, 0pt);
\coordinate (v4) at (0pt,24pt);
\coordinate (v5) at (-12pt, 24pt);
\coordinate (v6) at (12pt,24pt);
\draw[thick](v1)--(v2);
\draw[thick](v5)--(v6);
\draw[thick,decorate](v3)--(v4);
\node at (-18pt,0pt) {\scriptsize $\tilde{r}_1$};
\node at (18pt,0pt) {\scriptsize $\tilde{r}_2$};
\node at (-18pt,24pt) {\scriptsize $\tilde{r}_3$};
\node at (18pt,24pt) {\scriptsize $\tilde{r}_4$};
\node at (v1) [circle,fill,inner sep=1pt]{};
\node at (v2) [circle,fill,inner sep=1pt]{};
\node at (v5) [circle,fill,inner sep=1pt]{};
\node at (v6) [circle,fill,inner sep=1pt]{};
\end{tikzpicture}) + ~\mathcal{O}(\tilde{\phi}^6).
\end{eqnarray}
Higher ordered terms are contributed by all the connected Feynman diagrams with certain numbers of external field $\tilde{\phi}$ represented by black dots. This reproduces the previous result of the effective action as a series of the UV field\cite{moshe2003quantum,2006PhLB..632..571B,dupuis2021nonperturbative}.

\section{Scaling operators in the large N limit \label{sec:scaling_op}}

The exact fixed point effective action in \eq{eq:action_rescaling} 
is a function of $\tilde \phi_{\tilde r}$
 and $ X_{\tilde r}'$.
In order to understand the physical meaning of them,
let us deform the hopping amplitude at UV by
$-\frac{m^2}{2}\delta M_{ij} = 
\epsilon 
\Bigl[
\delta_{i,0} \delta_{j,\infty}
+\delta_{i,\infty} \delta_{j,0}
\Bigr]
+\epsilon' \delta_{ij}$.
The $\epsilon$-term is an infinite-range hopping,
which is equivalent to
inserting a pair of fundamental fields :
one at the origin
and the other at infinity.
$\epsilon'$ is a uniform mass deformation.
Under these variations,
${\bf M}$ and ${\bf X}$ are varied.
However, only the variation of ${\bf M}$ 
contributes to the change in the renormalized action to the linear order in 
$\epsilon$ and $\epsilon'$ 
because the action is stationary with respect to ${\bf X}$ at the saddle-point. The variation of the action is studied in Appendix \ref{app:scaling dimension}.
In the large  $z^\ast$ limit,
it becomes 
\begin{eqnarray}
\delta S_1^{z^\ast}
&=& -2\epsilon ~
e^{-2 \Delta_\phi z^\ast}
\phi^S_{ 0 }
\phi^S_{ \infty }  - \epsilon' 
e^{(D-\Delta_X)z^\ast}
\frac{m^2 N}{2 \lambda}
\int d^D\tilde r ~ 
{X}'_{\tilde r},
\label{eq:perturbation}
\end{eqnarray}
where 
\begin{eqnarray}
\phi^S_{\tilde r} 
&\equiv& \int d \tilde r' ~ 
\left[
\tilde{\bf T}+{\bf X}'
\right]^{-1}_{\tilde{r}\tilde{r}'}  
\tilde \phi_{\tilde r'}.
\label{eq:phi} \\
X'_{\tilde{r}}&=&x_2
+\frac{m^2}{N}\frac{\left[\frac{1}{ -\tilde{\nabla}^2/m^2 +\tilde{\Sigma}}\tilde{\phi}_{\tilde{r}}\right]^2 }{\frac{\Gamma (2-\frac{D}{2})}{(4\pi)^{D/2}}  \int_0^1 du \left[-u(1-u)\tilde{\nabla}^2/m^2 +\tilde{\Sigma}\right]^{\frac{D}{2}-2}} +\dots.
%\nonumber\\
\label{eq:x}
\end{eqnarray}
$\phi^S_{\tilde{r}}$ is an operator that has the same quantum number as $\phi$,
and is made of 
$\tilde \phi$ and $ X'$
with smearing at length scale
$\tilde \Sigma^{-1/2}$ 
in the scaled variable.
$X'_{\tilde{r}}$ is an 
$O(N)$-singlet composite operator which involves
infinitely many $\tilde \phi$'s and derivatives
(see Appendix \ref{app:solution_X_phi} for the expression).
\eq{eq:perturbation} shows that
$\phi^S_{\tilde r}$ and ${ X}'_{\tilde r}$ corresponds to the scaling operators whose forms are invariant under the coarse graining and dilatation.
$\phi^S_{\tilde r}$ 
and ${ X}'_{\tilde r}$ 
have scaling dimensions $\Delta_\phi=\frac{D-2}{2}$
and $\Delta_X=2$, respectively.
They are the leading scaling operators in the fundamental and singlet representations of the  $O(N)$ group, respectively.
We note that  
\eq{eq:perturbation} 
is obtained by taking the large $z^*$ limit of the effective action in the presence of the insertion of two UV operators.
At a finite $z^*$, 
\eq{eq:perturbation} 
is 
 corrected by extra terms that are further suppressed by higher powers of $e^{-z^*}$.
Since the scaling operators are 
 eigen-operators whose forms do not change under the coarse graining and dilatation, 
those corrections only contribute to the sub-leading scaling operators with larger scaling dimensions. 
The expressions for the leading scaling operators in Eqs. \eqref{eq:phi} and \eqref{eq:x} are exact in the large $N$ limit.

The first two terms in 
\eq{eq:action_rescaling}
can be written as
$ \frac{m^2}{2} 
\int d \tilde r d \tilde r' ~
{\phi}^S_{\tilde{r}} 
\left[ 
-(\tilde{\bf T}+{\bf X}' )
+(\tilde{\bf T}+{\bf X}' )^2
\right]_{\tilde{r}\tilde{r}'}
{\phi}^S_{\tilde{r}'}$,
and the entire fixed point action in 
\eq{eq:action_rescaling}
becomes a function of 
the two scaling operators only. 
From \eq{eq:saddle_point_rescaling}, 
 one can further find the relation between the two leading scaling operators,
\begin{eqnarray}
\frac{m^2}{N}
{\phi}_{\tilde{r}}^S \times {\phi}_{\tilde{r}}^S 
 + \left[\tilde{\bf T}+{\bf X}'\right]^{-1}_{\tilde{r}\tilde{r}}
-\left[
\tilde{\bf T} 
+  x_2 I 
\right]^{-1}
_{\tilde{r}\tilde{r}} 
= 0.  \nonumber
\end{eqnarray}
This gives the exact operator product expansion of two 
 $O(N)$ vector fields in terms of the singlet scaling operator and its descendants.

\section{Physical observables \label{sec:correlation_function}}

The effective action determines all $n$-point functions of the theory in the scaling limit. 
This follows from the fact that the full generating function 
\begin{eqnarray}
W[J] 
&=& -\ln \int 
{\mathcal D} \phi
~ e^{
- \frac{ m^2  }{2 }
\sum_{i}
\phi_{i}^2 
- S_1[ \phi ]
+ \sum_i J_i \phi_{i}
}
\end{eqnarray}
can be obtained from the scale dependent generating function
\begin{eqnarray}
W^z[J] 
&=& -\ln \int 
{\mathcal D} \phi
~ e^{
- \frac{ m^2_z  }{2 }
\sum_{i}
\phi_{i}^2 
- S_1[ \phi ]
+ \sum_i J_i \phi_{i}
} 
\nonumber\\
&=& 
-\ln \int 
\mathcal D \phi
~ e^{
- \frac{ \sum_{i}
J_{i}^2 }{2 m^2_z }
 -\frac{m_z^2}{2}\sum_i \phi_i^2
- S_1[ \phi + J/m_z^2]
},
\end{eqnarray}
by taking the scaling limit, where $m_z=\frac{m}{\sqrt{1-e^{-2z}}}$. Thus, it is related to the effective action 
\eq{eq:action_rescaling}
through
\begin{eqnarray}
W[J] = \lim_{z \rightarrow \infty} 
\left\{
- \frac{1}{2 m_z^2 }
\sum_{i}
J_{i}^2 
+ S_1[e^z J/m_z^2] \right\},
\label{eq:WtoS1}
\end{eqnarray}
Then, n-point functions can be obtained by taking derivatives of $W$ with respect to $J$. For example, the 2-point correlation function is given by $G_2^{ab}[r_1,r_2] =\langle \phi_{r_1}^a\phi_{r_2}^b\rangle = -\frac{\delta^2 W}{\delta J_{r_1}^a \delta J_{r_2}^b}$ where $a$ and $b$ are O(N) indices. Using Eq.~(\ref{eq:Stot}), we can get $G_2^{ab}[r_1,r_2] = \frac{\delta_{ab}}{m^2}[{\bf K}+x I]^{-1}_{r_1,r_2}$ where $x$ is the constant part of the saddle point solution of ${\bf X}$. Similarly, we can also compute 4-point function of $\phi$ (Appendix \ref{app:4pt}) which is given by
\begin{eqnarray}
&&G_4^{abcd}[r_1,r_2,r_3,r_4]=\frac{2}{Nm^4} \sum_{rr'} \Big(\delta_{ab}\delta_{cd}
\begin{tikzpicture}[baseline={([yshift=-4pt]current bounding box.center)}]
\tikzset{decoration={snake,amplitude=.4mm,segment length=1mm, post length=0mm,pre length=0mm}}
\coordinate (v4) at (0pt,12pt);
\coordinate (v3) at (20pt,12pt);
\coordinate (v2) at (0pt,-12pt);
\coordinate (v1) at (20pt,-12pt);
\coordinate (v5) at (10pt, -6pt);
\coordinate (v6) at (10pt,6pt);
\draw[thick](v1)--(v5);
\draw[thick](v2)--(v5);
\draw[thick](v3)--(v6);
\draw[thick](v4)--(v6);
\draw[thick,decorate](v5)--(v6);
\node at (-2pt,15pt) {\scriptsize $r_3$};
\node at (-2pt,-15pt) {\scriptsize $r_1$};
\node at (22pt,15pt) {\scriptsize $r_4$};
\node at (22pt,-15pt) {\scriptsize $r_2$};
\node at (5pt,-4pt) {\scriptsize $r'$};
\node at (5pt,4pt) {\scriptsize $r$};
\end{tikzpicture}
+\delta_{ac}\delta_{bd}
\begin{tikzpicture}[baseline={([yshift=-4pt]current bounding box.center)}]
\tikzset{decoration={snake,amplitude=.4mm,segment length=1mm, post length=0mm,pre length=0mm}}
\coordinate (v4) at (0pt,12pt);
\coordinate (v3) at (20pt,12pt);
\coordinate (v2) at (0pt,-12pt);
\coordinate (v1) at (20pt,-12pt);
\coordinate (v5) at (10pt, -6pt);
\coordinate (v6) at (10pt,6pt);
\draw[thick](v1)--(v5);
\draw[thick](v2)--(v5);
\draw[thick](v3)--(v6);
\draw[thick](v4)--(v6);
\draw[thick,decorate](v5)--(v6);
\node at (-2pt,15pt) {\scriptsize $r_2$};
\node at (-2pt,-15pt) {\scriptsize $r_1$};
\node at (22pt,15pt) {\scriptsize $r_4$};
\node at (22pt,-15pt) {\scriptsize $r_3$};
\node at (5pt,-4pt) {\scriptsize $r'$};
\node at (5pt,4pt) {\scriptsize $r$};
\end{tikzpicture}
+\delta_{ad}\delta_{bc}
\begin{tikzpicture}[baseline={([yshift=-4pt]current bounding box.center)}]
\tikzset{decoration={snake,amplitude=.4mm,segment length=1mm, post length=0mm,pre length=0mm}}
\coordinate (v4) at (0pt,12pt);
\coordinate (v3) at (20pt,12pt);
\coordinate (v2) at (0pt,-12pt);
\coordinate (v1) at (20pt,-12pt);
\coordinate (v5) at (10pt, -6pt);
\coordinate (v6) at (10pt,6pt);
\draw[thick](v1)--(v5);
\draw[thick](v2)--(v5);
\draw[thick](v3)--(v6);
\draw[thick](v4)--(v6);
\draw[thick,decorate](v5)--(v6);
\node at (-2pt,15pt) {\scriptsize $r_2$};
\node at (-2pt,-15pt) {\scriptsize $r_1$};
\node at (22pt,15pt) {\scriptsize $r_3$};
\node at (22pt,-15pt) {\scriptsize $r_4$};
\node at (5pt,-4pt) {\scriptsize $r'$};
\node at (5pt,4pt) {\scriptsize $r$};
\end{tikzpicture}
\Big)\nonumber
\end{eqnarray}
where 
\begin{eqnarray}
\begin{tikzpicture}[baseline={([yshift=-4pt]current bounding box.center)}]
\tikzset{decoration={snake,amplitude=.4mm,segment length=1mm, post length=0mm,pre length=0mm}}
\coordinate (v4) at (0pt,12pt);
\coordinate (v3) at (20pt,12pt);
\coordinate (v2) at (0pt,-12pt);
\coordinate (v1) at (20pt,-12pt);
\coordinate (v5) at (10pt, -6pt);
\coordinate (v6) at (10pt,6pt);
\draw[thick](v1)--(v5);
\draw[thick](v2)--(v5);
\draw[thick](v3)--(v6);
\draw[thick](v4)--(v6);
\draw[thick,decorate](v5)--(v6);
\node at (-2pt,15pt) {\scriptsize $r_3$};
\node at (-2pt,-15pt) {\scriptsize $r_1$};
\node at (22pt,15pt) {\scriptsize $r_4$};
\node at (22pt,-15pt) {\scriptsize $r_2$};
\node at (5pt,-4pt) {\scriptsize $r'$};
\node at (5pt,4pt) {\scriptsize $r$};
\end{tikzpicture} &=&-({\bf K}+x I)^{-1}_{r_3,r}({\bf K}+x I)^{-1}_{r,r_4}\mathbb{L}^{-1}_{rr'}\nonumber\\
&\times&({\bf K}+x I)^{-1}_{r',r_1}({\bf K}+x I)^{-1}_{r_2,r'}
\end{eqnarray}
and $\mathbb{L}$ is a matrix with elements $\mathbb{L}_{rr'}=[({\bf K}+x I)^{-1}_{rr'}]^2+\frac{m^4}{4\lambda}\delta_{rr'}$. General n-point function can be obtained in the same way.

\section{1/N corrections}

In the large $N$ limit, 
the integration of the collective variable in 
\eq{eq:S1zast0}
has been evaluated through
the saddle-point approximation.
For a finite $N$, one has to include fluctuations of the collective fields.
Writing fluctuations around the saddle point as
${\bf X} =\bar{\bf X} +\delta {\bf X}$ 
with $\delta {\bf X}_{ij}=\delta X_i \delta_{ij}$,
we express 
the effective action as 
\begin{eqnarray}
    |S_1^{z^{\ast}} \rangle &=& 
    \int \mathcal{D}\phi \mathcal{D}\delta X_{i}~\exp \Big\{ \frac{Nm^4\sum_i (\bar{X}_{i}+\delta X_i)^2}{16\lambda}  - \frac{N}{2}\textrm{tr}  \log \left[ (1-e^{-2z^\ast})({\bf K}+\bar{\bf X}+\delta {\bf X}) \right]  \nonumber\\
    &-&\frac{m^2}{2(e^{2z^\ast}-1)}\sum_{ij}\Big(I-\frac{e^{2z^\ast}}{(e^{2z^\ast}-1)}[{\bf K}+\bar{\bf X} +\delta {\bf X}]^{-1}\Big)_{ij}\phi_i\phi_j \Big\}|\phi\rangle  \nonumber\\
    & = & 
    \int \mathcal{D}\phi \mathcal{D}\delta X_{i} \exp\Big\{ -N\bar{S}[\phi,\bar{X}]-
    \Delta S[\phi,\bar{X},\delta X] 
    \Big\} |\phi\rangle,
\label{eq:25}
\end{eqnarray}
where $\Delta S$ 
 is the action for the fluctuating field, 
\begin{eqnarray}
    \Delta S [\phi,\bar{X},\delta X] &=& 
    \frac{N}{2}
    \left[
    \sum_{ij}
    G_{ij}^{-1}[\phi,\bar{X}] 
    \delta X_i 
    \delta X_j +
    \sum_{ijkl}\Gamma_{ijkl}[\phi,\bar{X}]\delta X_i\delta X_j\delta X_k\delta X_l 
    +\mathcal{O}[(\delta X)^6] \right].
    \nonumber \\
\end{eqnarray}
Here, $G_{ij}[\phi,\bar X]$ and $\Gamma_{ijkl}[\phi,\bar X]$ 
are the $\phi$-dependent propagator and quartic vertex 
for $\delta X$,
\begin{eqnarray}
    G^{-1}_{ij}[\phi,\bar{X}] &=& -\frac{m^4}{8\lambda} \delta_{ij} -\frac{1}{2}\Big( \frac{e^{2z^\ast}}{e^{2z^\ast}-1}\left[ {\bf K}+\bar{\bf X}\right]^{-1}_{ij}\Big)^2 \nonumber\\
    &-&\frac{m^2 e^{2z^\ast}}{2N(e^{2z^\ast}-1)^2} \sum_{kl}[{\bf K}+\bar{\bf X}]^{-1}_{ki}[{\bf K}+\bar{\bf X}]^{-1}_{ij}[{\bf K}+\bar{\bf X}]^{-1}_{jl}\phi_k \phi_l,
    \nonumber\\
    \Gamma_{ijkl} [\phi,\delta X] &=& -\sum_{a} [{\bf K}+\bar{\bf X}]^{-1}_{ai}[{\bf K}+\bar{\bf X}]^{-1}_{ij}[{\bf K}+\bar{\bf X}]^{-1}_{jk}[{\bf K}+\bar{\bf X}]^{-1}_{kl} \nonumber\\
    &\times&\left[\frac{1}{4}\Big(\frac{e^{2z^\ast}}{e^{2z^\ast}-1}\Big)^4 \delta_{a,l}+ \frac{m^2 e^{2z^\ast}}{2N(e^{2z^\ast}-1)^2}\sum_b [{\bf K}+\bar{\bf X}]^{-1}_{lb} \phi_a \phi_b\right].
\end{eqnarray}
The first two $1/N$ corrections
 to the effective action are obtained by integrating over $\delta X$ up to the quartic order in $\delta X$,
\begin{eqnarray}
    |S_1^{z^{\ast}} \rangle &\approx& \int \mathcal{D}\phi  \exp\Big\{ -N\bar{S}[\phi]-\delta S[\phi] \Big\} |\phi\rangle
\end{eqnarray}
where 
\begin{eqnarray}
    \delta S[\phi] &=& -\log \sqrt{\frac{(2\pi)^{L^2}}{N\textrm{det} {\bf G}^{-1}[\phi,\bar{X}] }} \left[1 + \frac{N}{2}\sum_{abcd}\Gamma_{abcd}[\phi,\bar{X}] \langle \delta X_a\delta X_b\delta X_c\delta X_d \rangle \right] \nonumber\\
    &=& \frac{1}{2}\Big(\log \left[\textrm{det} {\bf G}^{-1} \right] +\textrm{const}\Big)\left[1 + \frac{1}{2N}\sum_{abcd}\Gamma_{abcd} \Big(G_{ab}G_{cd} +G_{ac}G_{bd}+G_{ad}G_{bc}\Big)\right].
    \label{eq:DeltaS}
\end{eqnarray}
Higher order corrections can be similarly obtained.

The fluctuating collective variable $\delta X$ makes the hopping field $t_{ij}$ dynamical in the bulk.
Since the emergent geometry that the low-energy field is subject to at scale $z$ is controlled by the hopping field at that scale,
the fluctuating hopping field makes the bulk spacetime geometry dynamical\cite{lee2014quantum,lee2016horizon}.
However, the nature of the dynamical spacetime 
 that emerges 
 from \eq{eq:25}
 is rather special because
the collective field 
 in the bulk fluctuates  only through fluctuations of $\delta X$ at $z=0$
in \eq{eq:S1zast0}. 
For a configuration of the hopping field at the UV boundary, 
the hopping field in $z>0$ is completely fixed\footnote{
Inside the bulk, $p_{ij}$ acts as a Lagrangian multiplier that suppresses the fluctuations of $t_{ij}$\cite{dolan}.}.
This peculiarity arises due to the fact that in the vector model the theory that only includes the single-trace operators is free\cite{lee2014quantum}.

There is a way to obtain an 
 alternative bulk theory 
 in which the hopping field exhibit non-trivial fluctuations even in the bulk.
This can be achieved by including a quartic interaction in the reference action as\cite{lee2016horizon}
\begin{eqnarray}
     S'_{ref}   &=& \frac{m^2}{2}\sum_i \phi^2 +\frac{\lambda}{N}\sum_i (\phi_i^2)^2,  \nonumber\\
     S'_{1} &=& \frac{m^2}{2}\sum_{ij}M_{ij}\phi_i \phi_j.
\end{eqnarray}
Here, the quartic interaction is moved from $S_1$ to $S_{ref}$
and the partition function 
 $Z=\langle S_{ref} |S_1\rangle=\langle S'_{ref} |S'_1\rangle $ is unchanged. 
The RG Hamiltonian that leaves the new reference action invariant is obtained 
 from \eq{eq:RGH} through a similarity transformation,
\begin{eqnarray}
    \hat{H}' &=& e^{\frac{\lambda}{N}\sum_i (\phi_i^2)^2}\hat{H} e^{-\frac{\lambda}{N}\sum_i (\phi_i^2)^2} \nonumber\\
    &=&\sum_i \left[i  {\phi}_i {\pi}_i- \frac{4\lambda}{N}( {\phi}^2_i)^2+\frac{1}{m^2} {\pi}^2_i +\frac{i}{m^2}\frac{8\lambda}{N} {\phi}^2_i {\phi}_i {\pi}_i+\frac{4\lambda}{m^2}(1+\frac{2}{N}) {\phi}^2_i-\frac{16}{m^2}\frac{\lambda^2}{N^2}( {\phi}^2_i)^3\right].
\end{eqnarray}
Accordingly, the bulk action becomes
\begin{eqnarray}
    S'_{bulk}[z^\ast] &=& \int_0^{z^\ast} dz \Big[ i \sum_{kl}p_{kl}^z \partial_z t_{kl}^z -\frac{2i}{m^2}\sum_k t_{kk}^z +2i \sum_{kl}t^{z}_{kl}p^z_{kl} +\frac{4}{m^2}\sum_{ijk}t^z_{ik}t^z_{kj}p^z_{ij} \nonumber\\
    &+&\sum_k\Big(\frac{4\lambda (1+\frac{2}{N})}{m^2} p_{kk}^z-\frac{16\lambda^2}{m^2} (p_{kk}^z)^3-4\lambda (p_{kk}^z)^2\Big)+\frac{16i\lambda}{m^2}\sum_{kl}t^z_{kl}p_{ll}^zp_{kl}^z\Big].
\end{eqnarray}
In this alternative but exactly equivalent formulation,
the bulk theory includes 
 the hopping field that is genuinely dynamical.
The UV action is now linear in $p_{ij}$ :
$S'_{UV}=\sum_{ij}(it^0_{ij}+\frac{m^2}{2}M_{ij})p^0_{ij}$.
Integration over $p^0_{ij}$ at the UV boundary 
 imposes the Dirichlet boundary condition for the dynamical source at $z=0$,
 $t^0_{ij} = i \frac{m^2}{2}M_{ij}$.
 The partition function is given by the functional integration of the dynamcal source in the bulk,
\begin{eqnarray}
\left.    |S^{z^\ast}\rangle = \int \mathcal{D}\phi 
\left[ \int \mathfrak{D}t^{z>0} \mathfrak{D}p^{z>0} ~e^{-NS'_{bulk}[t,p]+i\sum_{ij}t_{ij}\phi_i\phi_j}  
\right|_{
 t^0_{ij} = i \frac{m^2}{2}M_{ij}
} \right]
|\phi\rangle.
\end{eqnarray}
In the large $N$ limit, this expression is reduced to 
\begin{eqnarray}
    |S^{z^\ast}\rangle \approx \int \mathcal{D}\phi e^{-NS'_{bulk}[\bar{t},\bar{p},z^\ast]+i\sum_{ij}\bar{t}_{ij}^{z^\ast}\phi_i\phi_j}  |\phi\rangle 
\end{eqnarray}
with fields $t$ and $p$ replaced by $\bar{t}$ and $\bar{p}$ as a saddle point solution of equations\cite{lee2016horizon}
\begin{eqnarray}
    i\partial_z p_{ij}^z &=& -\frac{2i}{m^2}\delta_{ij} + 2i p_{ij}^z+\frac{4}{m^2} \sum_k (t_{jk}^z p_{ik}^z+ t_{ki}^z p_{kj}^z) +\frac{8i\lambda }{m^2} (p^z_{ii}+p^z_{jj})p^z_{ij}, \nonumber\\
    i \partial_z t_{ij}^z &=& -2i t_{ij}^z - \frac{4}{m^2} \sum_{k}t_{ki}^z t_{kj}^z -\frac{4\lambda}{m^2}\delta_{ij}+8\lambda\delta_{ij}\Big(\frac{6\lambda}{m^2} p_{ii}^z+1 \Big)p_{ii}^z\nonumber\\
    &-&\frac{8i\lambda}{m^2} \delta_{ij}\sum_k(t^z_{ki}p^z_{ki}+t^z_{ik}p^z_{ik}) -\frac{8i\lambda}{m^2} t^z_{ij}(p^z_{jj}+p^z_{ii}).
\end{eqnarray}

The $1/N$ corrections can be incorporated by including fluctuations of the collective fields as
\begin{eqnarray}
    |S_1^{z^\ast} \rangle =\int \mathcal{D}\phi \mathfrak{D} \delta t^{z>0} \mathfrak{D} \delta p^{z>0}  e^{-NS'_{bulk}[\bar{t}+\delta t,\bar{p}+\delta p,z^\ast]+i\sum_{ij}\bar{t}_{ij}^{z^\ast}\phi_i\phi_j}  |\phi\rangle,
    \end{eqnarray}
where
\begin{eqnarray}
    S'_{bulk}[\bar{t}+\delta t,\bar{p}+\delta p,z^\ast] &=& S'_{bulk} [\bar{t},\bar{p},z^\ast] +\int_0^{z^\ast} dz \Big( i \sum_{kl} \delta p^z_{kl} \partial_z \delta t^z_{kl} +2i \sum_{kl}\delta t^z_{kl}\delta p^z_{kl} +\frac{4}{m^2}\sum_{ijk}\delta t^z_{ik} \delta {t}^z_{kj} \bar{p}^z_{ij}\nonumber\\
    &+&\frac{4}{m^2}\sum_{ijk}( \delta t^z_{ik}  {t}^z_{kj} \delta {p}^z_{ij}+ t^z_{ik}  \delta {t}^z_{kj} \delta {p}^z_{ij} ) -4\lambda \sum_k \left[\frac{12 \lambda}{m^2}\bar{p}^z_{kk} +1 \right](\delta p_{kk})^2 \nonumber\\
    &+&\frac{16 i\lambda}{m^2}\sum_{kl} \left[\delta t^z_{kl}\delta {p}_{ll}^z\bar{p}^z_{kl} +\delta t^z_{kl}\bar{p}_{ll}^z\delta {p}^z_{kl}+ \bar{t}^z_{kl}\delta {p}_{ll}^z\delta {p}^z_{kl} \right]\Big)  +\mathcal{O}[(\delta t)^3,(\delta p)^3].
\end{eqnarray}
Integration over $\delta t$ and $\delta p$ gives the leading $1/N$ correction to the effective action as in
    \eq{eq:DeltaS}.

\section{Summary and discussion \label{sec:conclusion}}

The main result of the paper is \eq{eq:action_rescaling}, which is a closed form of the Wilsonian effective action for the vector O(N) model in the large $N$ limit. 
Two comments are in order.
First, the effective action evaluated at $z^\ast$ is local at length scales $r \gg e^{z^\ast}/m$ (equivalently, $\tilde r \gg 1/m$).
This is because the effective action 
in \eq{eq:action_rescaling}
is obtained with the IR cutoff $m e^{-z^\ast}$.
The full effective action obtained for $z^\ast =\infty$ is non-local at the critical point. 
Second, the form of the effective action depends on the RG scheme because the way IR cutoff is imposed depends on the scheme (i.e., the choice of $S_{ref}$).
Nonetheless, the effective action for the modes with 
$q \gg m e^{-z^\ast}$
($\tilde q \gg m$)
should be independent of RG scheme.
Furthermore, 
the effective action at large momenta takes non-local form as expected.

In conclusion,  we obtained the exact Wilsonian effective action of the interacting $O(N)$ vector model.
It takes a closed form of  a transcendental function  of the two leading scaling operators in the large N limit,
where one is in the fundamental representation of $O(N)$
and the other is the singlet.
It will of great interest to extend the present result to fermionic systems and theories with non-local/imaginary couplings\cite{fisher1972critical,fisher1978yang}.

\section*{Acknowledgement}
Research at Perimeter Institute is supported in part by the Government of Canada through the Department of Innovation, Science and Economic Development Canada and by the Province of Ontario through the Ministry of Colleges and Universities.
SL acknowledges the support of the Natural Sciences and Engineering Research Council of Canada.

\bibliography{ref}

\clearpage
\onecolumngrid

\setcounter{equation}{0}
\setcounter{page}{1}
% \renewcommand{\theequation}{S.\arabic{equation}}

% \begin{center}
% {\bf \large Supplementary Material of ``Exact effective action for the O(N) vector model in the large N limit"}
% \end{center}

\appendix

\section{RG scheme \label{app:RG_main}}
\subsection{RG Hamiltonian  \label{app:RG_Hamiltonian_O(N)}}

In this section, we review the derivation of the RG Hamiltonian in Eq.~(\ref{eq:RGH}),
and elaborate why the quantum evolution generated by the RG Hamiltonian can be understood as a coarse graining transformation\cite{lee2016horizon}.
Consider a theory of a scalar whose 
partition function is 
\begin{equation}
\begin{aligned}
Z=\int \mathcal{D}\phi  ~ e^{-S_{ref}-S_1},
\end{aligned}
\end{equation}
where 
$S_{ref} =\frac{1}{2}m^2 \sum_{i} \phi_{i}^2$
is the reference action chosen to be the insulating fixed point action,
and  $S_1$ includes the kinetic term and all interactions.
As is discussed in the main text, 
this partition function can be written as the overlap between two quantum states,
$Z= \langle  S_{ref} | S_1\rangle$,
where 
$| S_{ref}\rangle =\int \mathcal{D}\phi e^{-S_{ref}}
|\phi\rangle$ 
and
$|S_1\rangle =\int \mathcal{D}\phi e^{-S_1 [\phi]}|\phi\rangle$.

In the real-space RG \`a la Kadanoff, 
a block of sites is merged into a coarse grained site at each step.
This forces the RG step to be discrete.
To avoid this, we adopt the real space RG scheme that keeps the number of sites unchanged under coarse graining.
In this scheme,
the field at each site
is partially integrated out by an infinitesimal amount without removing the site.
To facilitate this, we introduce an auxiliary field $\Phi$ with mass $\mu$ to the action as
\begin{equation}
S' = \frac{1}{2}m^2\sum_i \phi_i^2 
+\frac{1}{2}\sum_{i} \mu^2 \Phi_{i}^2 + S_1[\phi].
\end{equation}
Now the physical field and the auxiliary field are rotated into low-energy mode  ($\phi'$) and high-energy mode ($\tilde \phi$) as
\begin{equation}
\phi_i = \phi'_i+\tilde{\phi}_i,\quad \Phi_i =A \phi'_i +B \tilde{\phi}_i,
\end{equation}
where
$A =\frac{m^2}{\tilde{\mu} \mu}$
and
$B =-\frac{\tilde{\mu}}{\mu}$
with 
$\tilde{\mu}=
\frac{m}{\sqrt{e^{2
dz} -1 }}\approx\frac{m}{\sqrt{2
dz}}$.
Through this basis transformation,
the low-energy mode acquires a new  mass, $m e^{dz}$. 
The increased mass suppresses fluctuations of the low-energy mode slightly.
The missing fluctuation has been transferred to the high-energy mode,
which has a heavy mass with order of $m/\sqrt{dz}$.
In order to restore the reference action for the low-energy mode,
the fields are scaled as
\begin{equation}
\phi'_i = e^{-
dz} \phi''_i, \quad 
\tilde{\phi}_i = e^{-
dz} \tilde{\phi}''_i.
\end{equation}
In the new basis, the action reads
\begin{equation}
\begin{aligned}
S'' &=
\frac{1}{2}m^2\sum_{i}   (\phi_{i}'')^2 + \frac{1}{2}\sum_{i} \tilde{\mu}^2 (\tilde{\phi}_{i}'')^2
+ S_1[e^{-
dz}(\phi''_i+\tilde{\phi}''_i)].
\end{aligned}
\end{equation}
Integrating out the heavy mode $\tilde \phi''$,
the action is renormalized by $\delta S_1$ which is given by
\begin{eqnarray}
e^{- S_1[\phi''_i]-\delta S_1[\phi''_i] } =\left[1- 
dz\phi''_i \frac{\partial}{\partial \phi''_i} + 
\frac{
dz}{m^2} 
\left(
\frac{\partial }{\partial \phi''_i}
\right)^2
\right]
e^{-S_1[ \phi'']}.
\end{eqnarray}
This is the real space version of the exact Polchinski RG equation\cite{POLCHINSKI1984exactRG}.
The renormalization of the deformation can be understood as a result from
a quantum evolution acting on the wavefunction $e^{-S_1}$ :
$
e^{-S_1[\phi]-\delta S_1[\phi]}
=
e^{-\hat Hdz}e^{-S_1[\phi]}
$,
where $\hat H$ is the RG Hamiltonian,
\begin{eqnarray}
\hat H =\sum_i\left[ i \phi_i \pi_i + \frac{1}{m^2}\pi_i^2\right]
\end{eqnarray}
with $\pi_i = -i\frac{\partial }{\partial \phi_i}$.
One can readily check 
$H^\dag |S_{ref}\rangle =0$.
It is straightforward to  generalize this to the case for the $N$-component scalar fields as is shown in Eq.~(\ref{eq:RGH}).

\subsection{
IR cutoff of the effective action
\label{app:EffectiveAction}}

The renormalized deformation at scale $z$
is given by 
$S_1^z = - \ln \langle \phi | S_1^z \rangle$, where
$|S_1^z \rangle = e^{-\hat H z} | S_1^0 \rangle$.
In this section, we show that this renormalized action indeed represents the effective action obtained with a $z$-dependent IR cutoff.
From the discussion in the previous subsection,
the renormalized action after one infinitesimal RG step can be written as
\bqa
e^{-S_1^{dz}[\phi^{(1)}]} 
= \int {\mathcal D} \tilde{\phi}^{(1)}
e^{
- \frac{ \tilde{\mu}^2  }{2} \sum_{i}  
(\tilde{\phi}_{i}^{(1)})^2 
- S_1[e^{- dz}(\phi^{(1)}+\tilde{\phi}^{(1)})]
}.
\eqa
After one more step of coarse graining, 
the renormalized action  becomes
\bqa
e^{-S_1^{2dz}[\phi^{(2)}]} 
= \int 
{\mathcal D} \tilde{\phi}^{(1)}
{\mathcal D} \tilde{\phi}^{(2)}
e^{
- \frac{ \tilde{\mu}^2  }{2} \sum_{i}  
\left[
(\tilde{\phi}_{i}^{(1)})^2 
+
(\tilde{\phi}_{i}^{(2)})^2 
\right]
- S_1[
e^{- 2dz}
\phi^{(2)}
+ e^{- 2 dz} \tilde{\phi}^{(2)}
+ e^{- dz} \tilde{\phi}^{(1)}
]
}.
\eqa
Repeating this for $n$ steps,
we obtain
\bqa
e^{-S_1^{ndz}[\phi^{(n)}] } 
= 
\int  \prod_k {\mathcal D} \tilde{\phi}^{(k)}
e^{
- \frac{ \tilde{\mu}^2  }{2}
\sum_{i}
\sum_{k=1}^n
(\tilde{\phi}_{i}^{(k)})^2 
- S_1[
e^{- n dz} \phi^{(n)}
+ \sum_k e^{- k dz} \tilde{\phi}^{(k)}
]
}.
\label{eq:S10}
\eqa
Defining
$\phi_{i,<} = e^{-n dz} \phi_i^{(n)}$
and
$\phi_{i,>} = \sum_k e^{- k dz} \tilde{\phi}_i^{(k)}$
with $n = z^\ast/dz$,
we obtain
\bqa
e^{-S_1^{z^\ast}[ e^{z^\ast} \phi_< ] } 
= \int 
{\mathcal D} \phi_>
~ e^{
- \frac{ m_{z^*}^2  }{2 
}
\sum_{i}
\phi_{i,>}^2 
- S_1[ \phi_< + \phi_> ]
},
\label{eq:S11}
\eqa
where
$m_{z^*}
= \frac{m}
{\sqrt{ 1-e^{-2z^\ast}}}
$.
To obtain \eq{eq:S11},
we insert
$ 
1
=
\int 
{\mathcal D} \phi_>  
{\mathcal D} \gamma 
~ e^{i \sum_i \gamma_i( 
\phi_{>,i}
-
 \sum_k e^{- k dz} \tilde{\phi}^{(k)}_i
 ) }
$
in the integrand of \eq{eq:S10},
replace 
$ \sum_k e^{- k dz} \tilde{\phi}^{(k)} $
 with 
 $\phi_>$
 in $S_1$,
and integrate over $\tilde{\phi}^{(k)}$
and $\gamma$.
We note that \eq{eq:S11} is the effective action of a background field $\phi_<$ 
for a theory whose mass is greater than the mass of the original theory 
by $\delta m^2 = m^2 \frac{e^{-2z}}{1-e^{-2z}}$.
If the original theory is at the critical point, 
the new theory defined at $z^\ast$ has an IR cutoff $\delta m^2$. 
The IR cutoff varies from infinity to zero 
as $z^\ast$ changes from zero to infinity.
The one particle irreducible effective action with the $z$-dependent IR cutoff, 
which satisfies the exact Wetterich  RG equation\cite{WETTERICH1993evolutionequation}, can be readily obtained from $S_1^z$ through the Legendre transformation
\cite{morris}.

% \subsection{From the effective action to the generating function \label{app:generating_func}}

% To see how physical observables are related to the effective action,
% we define the scale dependent generating function,
% \bqa
% W^z[J, \phi_<] 
% = -\ln \int 
% {\mathcal D} \phi_>
% ~ e^{
% - \frac{ m_z^2  }{2 }
% \sum_{i}
% \phi_{i,>}^2 
% - S_1[ \phi_< + \phi_> ]
% + \sum_i J_i \phi_{i,>}
% }.
% \eqa
% We note that 
% $W[J] \equiv \lim_{z \rightarrow \infty}
% W^z[J,0]
% $
% is the full generating function of the theory,
% and all $n$-point functions can be obtained from $W[J]$.
% By shifting the integration variable as
% $\phi_{i,>} \rightarrow
% \phi_{i,>}  + \frac{J_i}{m_z^2}
% $ and using
% \eq{eq:S11},
% we obtain
% \bqa
% W^z[J, 0] 
% &=& 
% -\ln \int 
% {\mathcal D} \phi_>
% ~ e^{
%  \frac{1}{2 m_z^2 }
% \sum_{i}
% J_{i}^2 
% - \frac{ m_z^2  }{2 }
% \sum_{i}
% \phi_{i,>}^2 
% - S_1[ \phi_> + J/m_z^2 ]
% } \nn
% &=&
% - \frac{1}{2 m_z^2 }
% \sum_{i}
% J_{i}^2 
% + S^z_1[e^z J/m_z^2].
% \label{eq:WtoS1}
% \eqa
% \eq{eq:WtoS1} shows that
% the Wilsonian effective action at scale $z$ gives the  
% generating function of the theory  with the $z$-dependent IR cutoff.

\section{Bulk action \label{app:bulk_theory}}

In this section, we derive the bulk action in Eq.~(\ref{eq:Sbulk}). 
We start with $|S_1^0\rangle$ that defines the deformation added to the insulating fixed point action at UV,
\begin{eqnarray}
|S_{1}^0 \rangle = \int \mathcal{D}\phi e^{-\left[ \frac{\lambda}{N}\sum_{i }  ( \phi_{i}^2)^2
+\frac{m^2}{2}\sum_{ij} M_{ij} \phi_{i}\phi_{j}\right]}|\phi\rangle.
\end{eqnarray}
This state can be
expanded as a linear superposition of  $|t\rangle$ at $z=0$ defined in Eq.~(\ref{eq:t}) as
\begin{eqnarray}
|S_{1}^0 \rangle &=& \int \mathcal{D}t^0_{ij}~\int \mathcal{D}p^0_{ij}~\int \mathcal{D}\phi~ e^{-i\sum_{ij} t^0_{ij}(Np^0_{ij}-\phi_i\phi_j) } e^{-\left[ 
\frac{m^2}{2}\sum_{ij} M_{ij} \phi_{i}\phi_{j}+ \frac{\lambda}{N}\sum_{i }  ( \phi_{i}^2)^2\right]}|\phi\rangle\nonumber\\
&=& \int \mathcal{D}t^0_{ij}~ \int \mathcal{D}p^0_{ij} ~e^{-N S_{UV}[t^0_{ij},p^0_{ij}]}|t^0\rangle, 
\end{eqnarray}
where $S_{UV}[t^0_{ij},p^0_{ij}]=\sum_{ij}(it^0_{ij}+\frac{m^2}{2}M_{ij}) p^0_{ij}+\lambda\sum_{i }   (p^0_{ii})^2$. 
The $O(N)$ symmetry 
guarantees that the wavefunction for
$|S_1^z \rangle = e^{-\hat H z} |S_{1}^0 \rangle $
is a function of the bi-linear $\phi_i \phi_j$,
and 
$|S_1^z \rangle $
can be also spanned by
$|t\rangle$ at $z$ as
\begin{eqnarray}
|S_1^z \rangle &=& \int \mathcal{D}t^z_{ij}~\int \mathcal{D}p^z_{ij}~\int \mathcal{D}\phi~ e^{-i\sum_{ij} t^z_{ij}(Np^z_{ij}-\phi_i\phi_j) } e^{-S^z_1 [\phi_i\phi_j]}|\phi\rangle\nonumber\\
&=& \int \mathcal{D}t^z_{ij}~ \int \mathcal{D}p^z_{ij} ~ e^{-N\sum_{ij}it^z_{ij} p^z_{ij}-S^z_1 [N p^z_{ij}]}|t^z\rangle,
\end{eqnarray}
where $S_1^z[\phi_i \phi_j] = - \ln \langle \phi | S_1^z \rangle$,
and the superscript $z$ for  $t$, $p$ and $|t\rangle$ denotes RG time.
Therefore, it is enough to understand the evolution of the basis state under the RG Hamiltonian.
The evolution from $z$ to $z+dz$ of the basis state is given by
\begin{eqnarray}
e^{-\hat{H}[\hat{\phi},\hat{\pi}] dz}|t^z\rangle &=& \int \mathcal{D}\phi\left[1-\sum_k\Big( \phi_k\frac{\partial}{\partial \phi_k}-\frac{1}{m^2}\frac{\partial^2}{\partial \phi_k^2}\Big) dz\right]e^{i\sum_{ij}t^z_{ij}\phi_i\phi_j }|\phi\rangle \nonumber\\
&=& \int \mathcal{D}\phi \exp\left[-2\Big(  -\frac{iN}{m^2}\sum_k  t^z_{kk}+i \sum_{kl} t^z_{kl}\phi_k\phi_l+\frac{2}{m^2} \sum_{kji} t^z_{ki}t^z_{kj}\phi_i\phi_j\Big) dz\right]e^{i\sum_{ij}t^z_{ij}\phi_i\phi_j }|\phi\rangle \nonumber\\
&=& \int \mathcal{D}t^{z+dz}_{ij} \mathcal{D}p^{z+dz}_{ij}  e^{- dz N {\cal L}_{bulk}
}|t^{z+dz}\rangle,
\label{eq:sp11}
\end{eqnarray}
where 
$\mathcal{L}_{bulk}$ 
is the bulk Lagrangian with
\begin{eqnarray}
{\cal L}_{bulk}= i\sum_{ij}p^{z+dz}_{ij}\partial_z t^{z+dz}_{ij}-
\frac{2i}{m^2}\sum_k   {t}^{z+dz}_{kk} +2i \sum_{kl}  {t}^{z+dz}_{kl} {p}^{z+dz}_{kl}+\frac{4}{m^2} \sum_{kji}  {t}^{z+dz}_{ki} {t}^{z+dz}_{kj} {p}^{z+dz}_{ij}.
\end{eqnarray}
From this,
one can write 
$|S_1^z \rangle $
as the path integration shown in  \eq{eq:S1z}.

\section{
Derivation of the saddle-point equation in Eq. (\ref {eq:eqforX0}) 
\label{app:fixed_point_t^0}}

In Eq.~(\ref{eq:Sbulk}), $p_{ij}$ acts as a Lagrange multiplier that enforces the constraint,
\begin{eqnarray}
\partial_z t_{ij}+2t_{ij}-\frac{4 i }{m^2}\sum_k t_{ki}t_{kj}  &=& 0 \label{eq:constraint_bulk}
\end{eqnarray}
in the bulk.
We treat $ t_{ij}$ as a matrix. 
The solution $i{\bf  t}^{z}[{\bf t}^0] = (i{\bf  t}^{0})  \left[   e^{2z} -\frac{2}{m^2} (e^{2z}-1)(i{\bf  t}^{0})\right]^{-1}$
can be written as a function of $z$ 
and ${\bf t}^0$.
The partition function becomes
\begin{eqnarray}
Z
&=&
\int \mathcal{D} t_{ij}^0 \mathcal{D}p_{ij}^0 
\langle S_{ref} | 
e^{-NS_{UV}[{\bf t}^{0},{\bf p}^{0}]
+
N
\frac{2i}{m^2}\sum_i 
\int_0^{z^\ast} dz 
t_{ii}^z[{\bf t}^0]} |t^{z^\ast}[{\bf t}^0]\rangle.
\end{eqnarray}
Next, we integrate over $p_{ij}^{0}$ at the UV boundary to obtain
\begin{eqnarray}
Z
&=& \int \mathcal{D}\phi   \int \mathcal{D}t_{ij}^{0}~e^{- S_{tot}
} \left[\prod_{i\neq j}
\delta\left(
t_{ij}^{0}
-i\frac{m^2}{2}M_{ij} \right)
\right],
\end{eqnarray}
where the total  effective action is given by
\begin{eqnarray}
S_{tot} [\phi, {\bf X}] &=& \frac{1}{2}m^2\sum_i \phi_i^2+ \frac{1}{2}m^2\sum_{ij} \left(({\bf X}+{\bf M}) \left[   (e^{2z^\ast }-1)({\bf X}+{\bf M}+I)+I \right]^{-1}\right)_{ij}\phi_{i}\phi_{j}\\
&+& N \sum_i \int_0^{z^\ast} dz\left(({\bf X}+{\bf M}) \left[  (e^{2z }-1)({\bf X}+{\bf M}+I)+I \right]^{-1} \right)_{ii}
-\frac{N m^4}{16 \lambda}\sum_{i } X_{i}^2. \nonumber 
\end{eqnarray}
Here, the diagonal element of $t_{ij}^0$ which is not fixed by the delta function is singled out as
$-\frac{2i {\bf t}^{0}}{m^2}= {\bf X} +{\bf M}$ 
and ${\bf X}_{ij}=X_i\delta_{ij}$ 
is a diagonal matrix.
Integrating over 
$t^0_{i\neq j}$
and
${\bf X}$ in the  large $z^\ast$ limit gives the 
the IR fixed point action. 
In the large N limit, we can use the saddle point approximation.
The variation of the first two terms with respect to ${\bf X}$ gives 
\begin{eqnarray}
&& \sum_{jk}\partial_{{\bf X}_{ii}} \left[({\bf X}+{\bf M})\Big( I  +  (e^{2z^\ast }-1)({\bf X}+{\bf M}+I)\Big)^{-1}\right]_{ij}\phi_i\phi_j \nonumber\\
&&= e^{2z^\ast}\sum_{jk}
\left[I  +  (e^{2z^\ast }-1)({\bf X}+{\bf M}+I)\right]^{-1}_{ij}\phi_j\phi_k\left[I  +  (e^{2z^\ast }-1)({\bf X}+{\bf M}+I)\right]^{-1}_{ki}
\end{eqnarray}
without summation of $i$ index,
and the third term gives
\begin{eqnarray}
&&\partial_{\bf X} {\textrm{Tr}} \log \left[(e^{2z^\ast }-1)({\bf X}+{\bf M}+I) + 1\right]
= (e^{2z^\ast}-1)\left[I+(e^{2z^\ast}-1)({\bf X}+{\bf M}+I)\right]^{-1}.
\end{eqnarray}
Collecting these two contributions,
we obtain the saddle-point equation in Eq.~(\ref{eq:eqforX0}).

\section{Saddle point solution}

In this section, we present the saddle-point solution for ${\bf X}$.
We will separately discuss 
$ x(z) $,
which is the $\phi$ independent part, and ${\bf X}-x(z)$ which is $\phi$ dependent.

\subsection{Field independent part} 
\label{app:solution_x}

The field independent part of ${\bf X}$ denoted as 
${x}$ satisfies Eq.~(\ref{eq:equation_x}),
\begin{eqnarray}
&&(e^{2z }-1)\left[(e^{2z  }-1)({x}I + {\bf M}+I) +I\right]^{-1}_{ii}- \frac{m^4 }{4 \lambda}   {x} =0.
\end{eqnarray}
Fourier transformation of the matrix can be implemented through a unitary transformation, and the LHS becomes
\begin{eqnarray}
&&\left[(e^{2z }-1)( xI+{\bf M}+I)+I\right]^{-1} 
=U\left[ (e^{2z }-1)( xI+ {\bf M}'+I)+I \right]^{-1} U^{-1},
\end{eqnarray}
where $U_{in}=\frac{1}{\sqrt{V}}e^{iQ_n r_i}$,
$V$ is the number of sites,
and $ {\bf M}'=U^{-1} {\bf M} U$ which is given by
${ M}'_{mn}=\sum_{ij}U^{-1}_{mi} { M}_{ij}U_{jn}=\frac{1}{V}\sum_{ij}e^{-iQ_m r_i} { M}_{ij}e^{i Q_n r_j}={ M}'_{Q_n}\delta_{mn}$.
Above, we use the fact that
$M_{ij}$ depends only on the separation between $i$ and $j$ in the presence of 
the translational invariance.
In the momentum space,
the equation for $ x$ becomes
\begin{eqnarray}
\frac{1}{V} \sum_{n}  \frac{1}{ x+\frac{e^{2z}}{e^{2z}-1}+  {M}'_{Q_n}} &=& \frac{m^4 }{4 \lambda }  x. 
\end{eqnarray}
Here we use the simple dispersion
$ M'_{Q}=a+ \frac{Q^2}{m^2}$
with a hard momentum cutoff $\Lambda$.
For $D>2$, the left hand side of the equation becomes
\begin{eqnarray}
\int^{\Lambda } \frac{d^D  Q}{(2\pi)^D} \frac{1}{ \frac{Q^2}{m^2}+  x + \frac{e^{2z}}{e^{2z}-1} + a } &=&
c_1 m^2 \Lambda^{D-2}+c_2  m^4
\left(
x+\frac{e^{2z}}{e^{2z}-1}+a
\right)
\Lambda^{D-4} +O(\Lambda^{D-6}),
\end{eqnarray}
where $c_i$ are constants.
Writing $x=x_0 + x_2 e^{-2z} + ..$, at the critical point, 
we obtain
\begin{eqnarray}
x_0 = 
c_1
\frac{4\lambda}{m^2}  \Lambda^{D-2},
~~
x_2 = 
\frac{c_2 \Lambda^{D-4}}{\frac{1}{4\lambda}-c_2 \Lambda^{D-4}}
\end{eqnarray}
to the leading order in $\Lambda$.

\subsection{Field dependent part \label{app:solution_X_phi}} 

\begin{figure}[h]
    \centering
    \includegraphics[width=.8\textwidth]{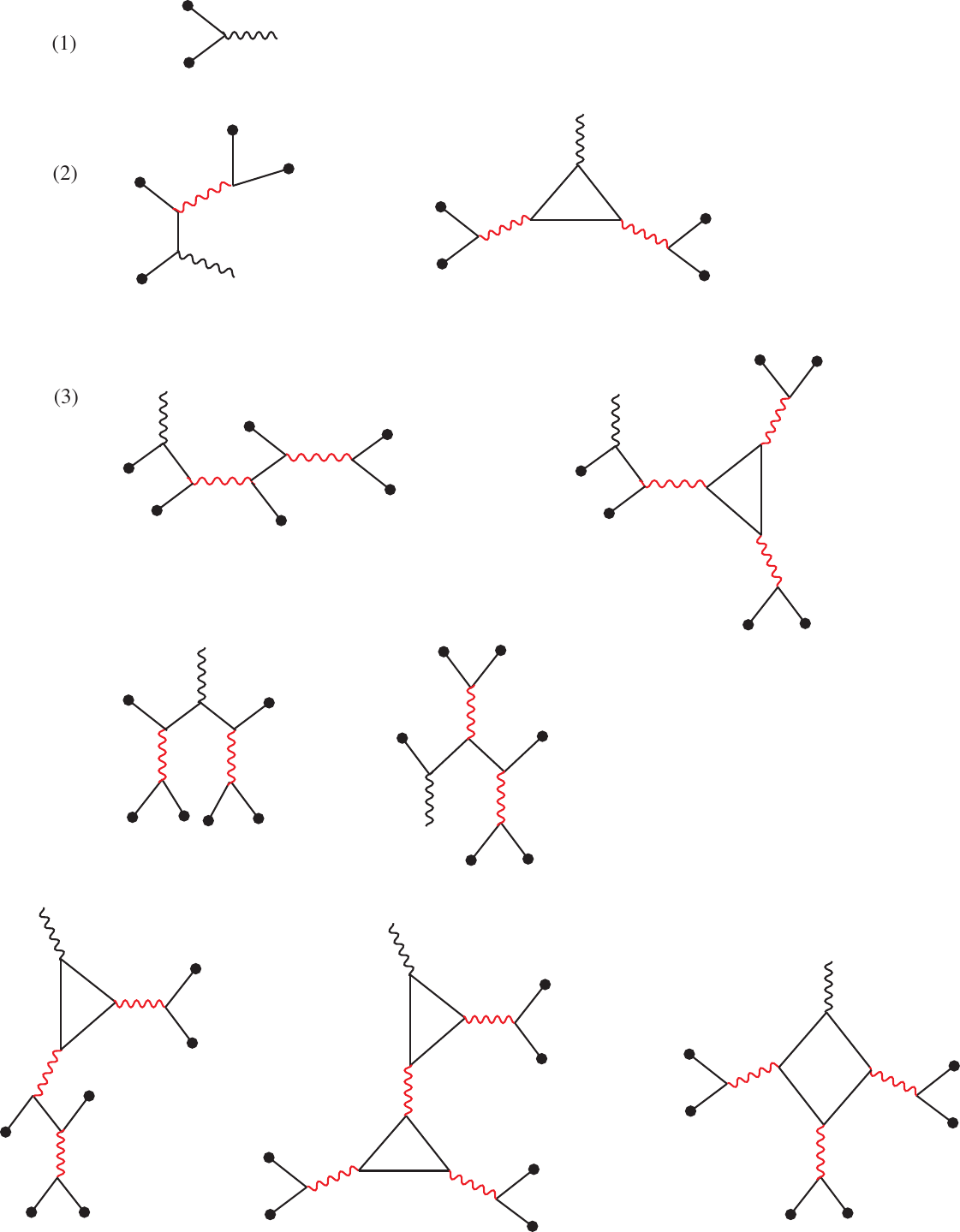}
    \caption{
    The Feynman diagrams representing the solution ${X}^{(k)}_{\tilde{r}}$ in Eq.~(\ref{eq:X123}). 
    Black dots denote $\tilde{\phi}_{\tilde{r}}$. 
The solid lines represent
the free propagators of $\phi$,
    $\tilde{\bf T}^{-1}_{\tilde{r}\tilde{r}'}$.
The wave lines are the 
propagators of the singlet field, 
$-\frac{1}{N}{\bf L}^{-1}_{\tilde{r}\tilde{r}'}$. 
There is an extra minus sign
for each propagator in red.
}
    \label{fig:sol_Feynman}
\end{figure}

In $D>4$, ${\bf X}'=x_2$.
Here we compute ${\bf X}'$ in $D<4$. 
At large $z^\ast$, 
the saddle point equation in Eq.~(\ref{eq:saddle_point_rescaling}) becomes
\begin{eqnarray}
&&   
\int d^D \tilde{r}_1 d^D \tilde{r}_2 ~\left[\tilde{\bf T}'+  {\bf X}'' \right]^{-1}_{\tilde{r}\tilde{r}_1}\tilde{\phi}'_{\tilde{r}_1}\tilde{\phi}'_{\tilde{r}_2}\left[\tilde{\bf T}'+  {\bf X}'' \right]^{-1}_{\tilde{r}_2\tilde{r}}+ \left[\tilde{\bf T}'+  {\bf X}'' \right]^{-1}_{\tilde{r}\tilde{r}} -   (\tilde{\bf T}')^{-1}_{\tilde{r}\tilde{r}}=0,
\label{eq:saddle_point_X'_d<4}
\end{eqnarray}
where 
$\tilde{\bf T}'=\tilde{\bf T}+x_2I$,
${\bf X}''={\bf X}'-x_2$
and
$\tilde{\phi}'_{\tilde{r}}=\frac{m}{\sqrt{N}}\tilde{\phi}_{\tilde{r}}$. 
Writing
$ {\bf X}''
= 
{\bf X}^{(1)}
+
{\bf X}^{(2)}
+
{\bf X}^{(3)}
+..$,
where 
$ {\bf X}^{(k)}$
is the $k$-th order term in
$\tilde{\phi}'_{\tilde{r}}\tilde{\phi}'_{\tilde{r}'}$,
we obtain an infinite series of recursion relations that determines
$ {\bf X}^{(k)}$ 
in terms of 
$ {\bf X}^{(n)}$ 
for $1 \leq n \leq k-1$.
The first few equations read  
\begin{eqnarray}
\left[(\tilde{\bf T}')^{-1}  {\bf X}^{(1)} (\tilde{\bf T}')^{-1} \right]_{\tilde{r}\tilde{r}}&=& \int d^D \tilde{r}_1 d^D \tilde{r}_2 (\tilde{\bf T}')^{-1}_{\tilde{r}\tilde{r}_1}(\tilde{\phi}'_{\tilde{r}_1} \tilde{\phi }'_{\tilde{r}_2})(\tilde{\bf T}')^{-1}_{\tilde{r}_2\tilde{r}}, \nonumber\\
\left[(\tilde{\bf T}')^{-1}  {\bf X}^{(2)} (\tilde{\bf T}')^{-1} \right]_{\tilde{r}\tilde{r}}&=& -2\int d^D \tilde{r}_1 d^D \tilde{r}_2 \left[ (\tilde{\bf T}')^{-1}  {\bf X}^{(1)}  (\tilde{\bf T}')^{-1}\right]_{\tilde{r}\tilde{r}_1}(\tilde{\phi}'_{\tilde{r}_1} \tilde{\phi }'_{\tilde{r}_2}) \left[ (\tilde{\bf T}')^{-1}\right]_{\tilde{r}_2\tilde{r}} \nonumber\\
&+&\left[ (\tilde{\bf T}')^{-1}  {\bf X}^{(1)} (\tilde{\bf T}')^{-1}   {\bf X}^{(1)} (\tilde{\bf T}')^{-1} \right]_{\tilde{r}\tilde{r}}, \nonumber\\
\left[(\tilde{\bf T}')^{-1}  {\bf X}^{(3)} (\tilde{\bf T}')^{-1} \right]_{\tilde{r}\tilde{r}}&=&-2\int d^D \tilde{r}_1 d^D \tilde{r}_2\left[(\tilde{\bf T}')^{-1}  {\bf X}^{(2)} (\tilde{\bf T}')^{-1}\right]_{\tilde{r}\tilde{r}_1}(\tilde{\phi}'_{\tilde{r}_1} \tilde{\phi }'_{\tilde{r}_2}) \left[ (\tilde{\bf T}')^{-1}\right]_{\tilde{r}_2\tilde{r}}\nonumber\\
&+&\int d^D \tilde{r}_1 d^D \tilde{r}_2\left[(\tilde{\bf T}')^{-1}  {\bf X}^{(1)} (\tilde{\bf T}')^{-1} \right]_{\tilde{r}\tilde{r}_1}(\tilde{\phi}'_{\tilde{r}_1} \tilde{\phi }'_{\tilde{r}_2})\left[(\tilde{\bf T}')^{-1}  {\bf X}^{(1)} (\tilde{\bf T}')^{-1}\right]_{\tilde{r}_2\tilde{r}}\nonumber\\
&+& 2\int d^D \tilde{r}_1 d^D \tilde{r}_2\left[(\tilde{\bf T}')^{-1}   {\bf X} ^{(1)}(\tilde{\bf T}')^{-1}  {\bf X}^{(1)}(\tilde{\bf T}')^{-1}\right]_{\tilde{r}\tilde{r}_1}(\tilde{\phi}'_{\tilde{r}_1} \tilde{\phi }'_{\tilde{r}_2}) \left[ (\tilde{\bf T}')^{-1}\right]_{\tilde{r}_2\tilde{r}}\nonumber\\&+&
2 \left[(\tilde{\bf T}')^{-1}  {\bf X}^{(2)} (\tilde{\bf T}')^{-1}   {\bf X}^{(1)} (\tilde{\bf T}')^{-1}\right]_{\tilde{r}\tilde{r}} \nonumber\\
&-& \left[(\tilde{\bf T}')^{-1}  {\bf X}^{(1)} (\tilde{\bf T}')^{-1}   {\bf X}^{(1)} (\tilde{\bf T}')^{-1}   {\bf X}^{(1)} (\tilde{\bf T}')^{-1}\right]_{\tilde{r}\tilde{r}}.
\label{eq:X123}
\end{eqnarray}
Let us solve ${\bf X}^{(1)}$ explicitly.
If we view $  X^{(1)}_{\tilde{r}}$ as the $\tilde{r}$-th element of a vector $\vec{X}^{(1)}$, 
the first equation 
in \eq{eq:X123}
is written as ${\bf L}\vec{X}^{(1)}=\vec{\Phi}$,
where 
\begin{eqnarray}
{\bf L}_{\tilde{r}\tilde{r}'} &=&  \left[(\tilde{\bf T}')^{-1}_{\tilde{r}\tilde{r}'}\right]^2 = \left[\int \frac{d^D \tilde{Q}}{(2\pi)^D} \frac{e^{i \tilde{Q}(\tilde{r}-\tilde{r}')}}{\tilde{Q}^2/m^2+\tilde{\Sigma}}\right]^2,  \nonumber\\
\vec{\Phi}_{\tilde{r}} &=& \left[ \int d^D \tilde{r}' (\tilde{\bf T}')^{-1}_{\tilde{r}\tilde{r}'} \tilde{\phi}'_{\tilde{r}'}\right]^2 =\left[ \int d^D \tilde{r}' \int \frac{d^D \tilde{Q}}{(2\pi)^D} \frac{e^{i \tilde{Q}(\tilde{r}-\tilde{r}')}}{\tilde{Q}^2/m^2+\tilde{\Sigma}} \tilde{\phi}'_{\tilde{r}'}\right]^2.
\end{eqnarray}
The solution can be obtained by inverting ${\bf L}$ as
\begin{eqnarray}
X_{\tilde{r}}^{(1)}
&=& \frac{m^2}{N}\frac{1 }{\frac{\Gamma (2-\frac{D}{2})}{(4\pi)^{D/2}}  \int_0^1 du \left[-u(1-u)\tilde{\nabla}^2/m^2 +\tilde{\Sigma}\right]^{\frac{D}{2}-2}}\left[\frac{1}{ -\tilde{\nabla}^2/m^2 +\tilde{\Sigma}}\tilde{\phi}_{\tilde{r}}\right]^2,
\end{eqnarray}
where the answer is written in terms of 
$\tilde{\phi}_{\tilde{r}}$. 
It is noted that 
$\frac{1}{N}{\bf L}^{-1}_{\tilde{r}\tilde{r}'}$ actually gives the propagator of singlet field in the large N limit while $\tilde{\bf T}^{-1}_{\tilde{r}\tilde{r}'}$ is the free propagator. 
One can use Feynman diagrams to represent ${X}^{(k)}_{\tilde{r}}$ as is shown in Fig.~\ref{fig:sol_Feynman}.

\section{ Leading scaling operators at the Wilson-Fisher fixed point \label{app:scaling dimension}}

The Wilson-Fisher fixed point has two relevant scaling operators.
In this section, we compute them in $D<4$.
In order to extract a scaling operator with a certain symmetry charge, 
we perturb the UV action with  an operator that has the same symmetry charge, 
and study the RG flow induced by it. 
The RG flow can be decomposed into eigen-modes each of which has a definite scaling dimension. 

\subsection{O(N) singlet\label{app:singlet}}

To extract the leading singlet operator,
let us perturb the system 
with a uniform mass :
$-\frac{m^2}{2}{\bf M}'_{ij}=-\frac{m^2}{2}{\bf M}^c_{ij}+{ \epsilon}'\delta_{ij}$,
where $-\frac{m^2}{2}{\bf M}^c$ is a hopping matrix that flows to the Wilson-Fisher fixed point. 
In the presence of the perturbation, 
the saddle point solution of  $ {\bf X}$ is modified to
$ {\bf X}+ \delta {\bf X}$. 
However, the variation of ${\bf X}$ does not contribute to the change of the effective action  to the linear order in $\epsilon'$ because 
the effective action is stationary with respect to ${\bf X}$ at the  saddle point. 
As a result, the variation of the effective action 
caused by the change in ${\bf M}$ becomes 
\begin{eqnarray}
\Delta_{\epsilon} S_{tot}
&=&-\frac{2}{m^2}{ \epsilon}' \sum_i
\partial_{M_{ii}}S_{tot} \nn
&=&  -\frac{2}{m^2}\epsilon'\int d^D r \left( \frac{e^{2z^\ast}}{(e^{2z^\ast}-1)^2}\int d^D r_1 d^D r_2~m^2 
\left[{\bf K}+ {\bf X}\right]^{-1}_{rr_1}(\phi_{r_1} \phi_{r_2})\left[{\bf K}+ {\bf X}\right]^{-1}_{r_2r}+N \left[{\bf K}+ {\bf X}\right]^{-1}_{rr} \right) \nonumber \\
&=&
- \epsilon' 
\frac{m^2 N}{2 \lambda}
\int d^D \tilde{r} 
\left[
e^{Dz^\ast}   x_0
+ e^{(D-2)z^\ast}
{X}'_{\tilde{r}}
\right].
\end{eqnarray}
From the second line to the last line, the saddle point equation in Eq.~(\ref{eq:eqforX0}) is used.
The final answer is expressed in terms of  the rescaled  coordinate,
$\tilde r = r e^{-z^\ast}$.
The first term is the correction to the identity operator with scaling dimension $0$.
It describes the change of the free energy.
The second term shows that
$ {X}_{\tilde{r}}'$ is the next leading singlet scaling operator with the scaling dimension $\Delta_{X}=2$.

\subsection{$O(N)$ non-singlet \label{app:vector}}

In order to extract the scaling operator in the fundamental representation of $O(N)$,
we add a non-local hopping term between the origin and $R$ 
in the $R \rightarrow \infty$ limit.
This is equivalent to inserting two fundamental fields far from each other.
For this, we consider a deformation given by
$-\frac{m^2}{2}{\bf M}'_{rr'}=-\frac{m^2}{2}{\bf M}^c_{rr'}+{ \epsilon}
\Bigl[
\delta(r)\delta(r'-R)
+\delta(r')\delta(r-R)
\Bigr]
$.
In the large $\tilde R$ limit,
the change in the IR action becomes
\begin{eqnarray}
\Delta_\epsilon S_{tot} 
&=& -\frac{4}{m^2}\epsilon \left( \frac{e^{2z^\ast}}{(e^{2z^\ast}-1)^2}
\frac{m^2}{2}
\int d^D r_1 d^D r_2 
\left[{\bf K}+ {\bf X}\right]^{-1}_{0 r_1}(\phi_{r_1} \phi_{r_2})\left[{\bf K}+ {\bf X}\right]^{-1}_{r_2 R}+
\frac{N}{2} \left[{\bf K}+ {\bf X}\right]^{-1}_{0 R} \right) \nonumber\\
&=& -2 \epsilon 
e^{(2-D)z^\ast} \int d^D \tilde{r}_1d^D \tilde{r}_2~  
\left[\tilde{\bf T}+ {\bf X}'\right]^{-1}_{0 \tilde{r}_1}(\tilde{\phi}_{\tilde{r}_1} \tilde{\phi}_{\tilde{r}_2})\left[\tilde{\bf T}+ {\bf X}'\right]^{-1}_{\tilde{r}_2\tilde{R}}.
\end{eqnarray}
In the last line, 
$\lim_{\tilde R \rightarrow \infty} \left[\tilde{\bf T}+ {\bf X}'\right]^{-1}_{0 \tilde R } = 0$ is 
used for which large $R$ limit is taken first 
before the large $z$ limit.
In this limit,
$\Delta_\epsilon S_{tot}$ can be viewed
as two local operators each of which inserted at the origin and the infinity, respectively.
This shows that 
$\phi^S_{\tilde r}=\int d^D \tilde{r}_1 [\tilde{\bf T}+ {\bf X}']^{-1}_{\tilde{r}\tilde{r}_1}\tilde{\phi}_{\tilde{r}_1}$ 
is the scaling operator in the fundamental representation of $O(N)$ with scaling dimension $\Delta_{\phi}=\frac{D-2}{2}$.

\section{Rescaling of spacetime and field \label{app:rescaling}}

In this appendix, 
we elaborate on the derivations of  
Eqs.~
(\ref{eq:action_rescaling}) and 
(\ref{eq:saddle_point_rescaling}) 
from 
Eqs.~(\ref{eq:Stot}) 
and ~(\ref{eq:eqforX0}).
For convenience, the lattice spacing is chosen to be $1$ at UV.
Matrix multiplications in real space are written as integrations as 
$
({\bf A} {\bf B})_{r_1 r_2}
=
\sum_r A_{r_1 r} B_{r r_2} = \int dr A_{r_1 r} B_{r r_2}$
in which
the identity matrix  becomes 
$I_{rr'}=\delta (r-r')$.
In order to make it manifest that the effective action is scale invariant at the critical point,
we use the rescaled momentum, coordinate and field defined by
$\tilde{Q}=Qe^{z^\ast}$,
$\tilde{r}=re^{-z^\ast}$
and 
$\tilde{\phi}_{\tilde{r}}=\phi_r e^{\frac{D}{2}z^\ast}$. 
Taking the large $z$ limit with fixed $\tilde Q$ is tantamount to zooming in toward the neighbourhood of $Q=0$ which contains the dynamical information on the universal long-distance physics.

The kernel in the first term of Eq.~ (\ref{eq:Stot}) 
is written as
\begin{eqnarray}
&&\left[(e^{2z^\ast}-1)({\bf K}+  {\bf X}) \right]^{-1}_{rr'} =\left[(e^{2z^\ast}-1)({\bf K}+ {x}_0I)+(e^{2z^\ast}-1)( {\bf X}- {x}_0I) \right]^{-1}_{rr'}.
\label{eq:S37}
\end{eqnarray}
Treating
\begin{eqnarray}
\tilde{\bf T}^{-1}_{\tilde{r}\tilde{r}'}
\equiv
e^{Dz^\ast}
\left[(e^{2z^\ast}-1)({\bf K}+ {x}_0I)\right]^{-1}_{rr'} 
&=&
\int^{\Lambda e^{z^\ast}} \frac{d^D \tilde{Q}}{(2\pi)^D} \frac{e^{i \tilde{Q}(\tilde{r}-\tilde{r}')}}{ \tilde{Q}^2/m^2+1}
\nonumber\\ 
\label{eq:rescaling_Tinv}
\end{eqnarray}
as the zero-th order term,
one can expand \eq{eq:S37} in powers of 
$  {\bf X}'_{\tilde r \tilde r'} \equiv  e^{2z^\ast}  \delta( \tilde r - \tilde r')  (   X_{r} -   x_0 ) $
as
\begin{eqnarray}
&&
e^{-Dz^\ast}
\Bigl\{
\tilde{\bf T}^{-1}_{\tilde{r}\tilde{r}'}
-\sum_{M_1 \geq 1}^\infty
(-1)^{M_1-1}\int d^D \tilde{R}~ \tilde{\bf T}^{-1}_{\tilde{r}\tilde{R}
}   {\bf X}'_{\tilde{R} \tilde{R}} \int d^D \tilde{R}_1~ \tilde{\bf T}^{-1}_{\tilde{R}\tilde{R}_1}   {\bf X}'_{\tilde{R}_1\tilde{R}_1} \int d^D \tilde{R}_2~\tilde{\bf T}^{-1}_{\tilde{R}_1\tilde{R}_2} \nonumber\\
&&\times \dots \times \int d^D \tilde{R}_{M_1-1} \tilde{\bf T}^{-1}_{\tilde{R}_{M_1-2}\tilde{R}_{M_1-1}}  {\bf X}'_{\tilde{R}_{M_1-1}\tilde{R}_{M_1-1}} \tilde{\bf T}^{-1}_{\tilde{R}_{M_1-1}\tilde{r}'} 
\Bigr\}
=e^{-Dz^\ast}\left[\tilde{\bf T}+ {\bf X}' \right]^{-1}_{\tilde{r}\tilde{r}'}.
\end{eqnarray}
Consequently,
\begin{eqnarray}
\left[(e^{2z^\ast}-1)({\bf K}+  {\bf X}) \right]^{-1}_{rr'}  &=& e^{-Dz^\ast} \left[\tilde{\bf T}+  {\bf X}' \right]^{-1}_{\tilde{r}\tilde{r}'}.
\label{eq:rescaling_T_X_combine}
\end{eqnarray}

The second term of Eq.~ (\ref{eq:Stot}) can be written as
\begin{eqnarray}
&&\sum_r \log \left[  (e^{2z^\ast}-1)({\bf K}+ {\bf X})  \right]_{rr} 
= \sum_r \log \left[ I + A \right],
\end{eqnarray}
where 
$A=  (e^{2z^\ast}-1)({\bf K}+ {\bf X}) -I$.
An expansion in powers of $A$ gives
\begin{eqnarray}
&&\sum_r \log \left[  
I + A \right]
= 
\sum_{M'\geq 1}\frac{(-1)^{M'-1}}{M'} 
\sum_{r_1, .., r_{M'}}
A_{r_1 r_2}
..
A_{r_{M'} r_{1}}.
\end{eqnarray}
In the rescaled coordinate,
the elements of $A$ becomes
$A_{rr'}
= e^{-D z^\ast}
\left[
(\tilde {\bf T} +   {\bf X}')_{\tilde r \tilde r'}
- \delta( \tilde r - \tilde r')
\right]
$
and $\sum_r = e^{D z^\ast} \int d^D\tilde r$.
From this, 
we obtain
\begin{eqnarray}
\sum_r \Big\{\log \left[ (e^{2z^\ast}-1)({\bf K}+ {\bf X}) \right]\Big\}_{rr}  =
\int d^D\tilde r
\Big\{\log \left[ \tilde{\bf T}+ {\bf X}' \right] \Big\}_{\tilde{r}\tilde{r}}.
\label{eq:rescaling_log}
\end{eqnarray}
Using Eq.~(\ref{eq:rescaling_Tinv}), Eq.~(\ref{eq:rescaling_T_X_combine})
and Eq.~(\ref{eq:rescaling_log}), we can obtain Eq.~(\ref{eq:action_rescaling}) and Eq.~(\ref{eq:saddle_point_rescaling}) from Eq.~(\ref{eq:Stot}) and Eq.~(\ref{eq:eqforX0}).

\section{Correlation functions from generating function \label{app:correlation_func}}

The $n$-point function can be
expressed as derivatives of 
the generating function,
\begin{eqnarray}
\langle \phi_{r_1}^{a_1}\phi_{r_2}^{a_2}\dots \phi_{r_n}^{a_n} \rangle =-\frac{\delta^n W[J]}{\delta J_{r_1}^{a_1} \delta J_{r_2}^{a_2} \dots \delta J_{r_n}^{a_n} }\Big|_{J=0},
\end{eqnarray}
where
the full generating function 
is given by the effective action
in the large $z$ limit
as is shown in Eq.~(\ref{eq:WtoS1}).
In this appendix, we explicitly compute the $2$-point function and the $4$-point function using the exact effective action
in \eq{eq:Stot}.

\subsection{2-pt correlation function \label{app:2pt}}

The generating function 
is a function of
$J_i^{a}$ and 
$X_k[J]$,
where $X_k[J]$
satisfies the saddle point equation 
in Eq.~(\ref{eq:eqforX0}). 
The derivative of $W$ with respect to a source is written as
$\frac{\delta W[J]}{\delta J_i^{a} }= \frac{\partial W}{\partial J_i^a}+\sum_k \frac{\partial W}{\partial X_k}\frac{\partial X_k}{\partial J_i^a}\Big|_{X=\bar{X}}$,
and the 2-pt correlation function can be expressed as
\begin{eqnarray}
G_2^{ab}[r_1,r_2] &=& - \frac{\delta^2 W}{\delta J_{r_1}^a \delta J_{r_2}^b  }\Big|_{J=0} = 
- \left[\frac{\partial }{\partial J_{r_1}^a}+\sum_r \frac{\partial X_r}{\partial J_{r_1}^a}\frac{\partial }{\partial X_r}\right]\left[\frac{\partial W}{\partial J_{r_2}^b}+\sum_{r'} \frac{\partial W}{\partial X_{r'}}\frac{\partial X_{r'}}{\partial J_{r_2}^b}\right]_{\substack{J=0 \\ \vspace{-1.5mm} \\ X=\bar{X}}}\nonumber\\
&=&- \Big[\frac{\partial^2 W}{ \partial J_{r_1}^a \partial J_{r_2}^b}+ \sum_r\frac{\partial^2 W}{\partial J_{r_1}^a \partial X_r}\frac{\partial X_r}{\partial J_{r_2}^b} +\sum_r \frac{\partial X_r}{\partial J_{r_1}^a} \frac{\partial^2 W}{\partial X_r \partial J_{r_2}^b} +\sum_{rr'}\frac{\partial^2 W}{\partial X_r \partial X_{r'}} \frac{\partial X_r}{\partial J_{r_1}^a}\frac{\partial X_{r'}}{\partial J_{r_2}^b} \nonumber\\
&+& \sum_r \frac{\partial W}{\partial X_r}\frac{\partial^2 X_r}{\partial J_{r_1}^a \partial J_{r_2}^b}\Big]_{\substack{J=0 \\ \vspace{-1.5mm} \\ X=\bar{X}}} \nonumber\\
&=&- \left[\frac{\partial^2 W}{ \partial J_{r_1}^a \partial J_{r_2}^b}+ \sum_r \frac{\partial W}{\partial X_r}\frac{\partial^2 X_r}{\partial J_{r_1}^a \partial J_{r_2}^b}\right]_{\substack{J=0 \\ \vspace{-1.5mm} \\ X=\bar{X}}}.
\end{eqnarray}
In the last equality, we used the fact that  both $W$ and $X$ are even functions of $J^a_{r}$.
Besides, we also have $\frac{\partial W}{\partial X_l}\Big|_{ X=\bar{X}}=0$ 
because $\bar{X}_l$ satisfies the saddle point equation of $W$. 
Thus, only the first term contributes to the 2-pt correlation of $\phi$, 
\begin{eqnarray}
 \frac{\partial^2 W}{ \partial J_{r_1}^a \partial J^b_{r_2}} &=&  -\frac{\delta_{ab}}{m_z^2}\delta_{r_1,r_2} +\frac{\partial^2 S_1^z [e^z J/m_z^2]}{\partial J_{r_1}^a \partial J_{r_2}^b} =-\frac{\delta_{ab} }{m^2} \left[{\bf K}+{\bf X}\right]^{-1}_{r_1,r_2}.
\end{eqnarray}
Setting $J^a_r=0$ and ${\bf X}=x(z) I$ at the critical point with $x(z)+a+1=x_2 e^{-2z}$, we obtain 
\begin{eqnarray}
G_2^{ab}[r_1,r_2] &=& \frac{\delta_{ab}}{m^2}\int \frac{d^D Q}{(2\pi)^D} \frac{e^{iQ\cdot (r_1-r_2)}}{\frac{Q^2}{m^2}+x_2e^{-2z}} \xrightarrow{z \rightarrow \infty} \frac{\delta_{ab}}{|r_1-r_2|^{D-2}}.
\end{eqnarray}
This is the free propagator of the fundamental field
as expected in the large $N$ limit.

\subsection{4-pt correlation function \label{app:4pt}}

The 3-point function is zero since it does not respect the O(N) symmetry. 
The 4-point function is expressed as
\begin{eqnarray}
G_4^{abcd}[r_1,r_2,r_3,r_4] &=& -
\frac{\delta^4 W}{\delta J^a_{r_1} \delta J^b_{r_2} \delta J^c_{r_3}  \delta J^d_{r_4} }\Big|_{J=0},
\end{eqnarray}
where
\begin{eqnarray}
\frac{\delta^4 W}{\delta J^a_{r_1} \delta J^b_{r_2} \delta J^c_{r_3}  \delta J^d_{r_4} } &=&  
\left[\frac{\partial }{\partial J_{r_4}^d}+\sum_r \frac{\partial X_r}{\partial J_{r_4}^d}\frac{\partial }{\partial X_r}\right]\left[\frac{\partial }{\partial J_{r_3}^c}+\sum_r \frac{\partial X_r}{\partial J_{r_3}^c}\frac{\partial }{\partial X_r}\right]\left[\frac{\partial }{\partial J_{r_2}^b}+\sum_r \frac{\partial X_r}{\partial J_{r_2}^b}\frac{\partial }{\partial X_r}\right]\nonumber\\
&\times&\left[\frac{\partial W}{\partial J_{r_1}^a}+\sum_r \frac{\partial W}{\partial X_r}\frac{\partial X_r}{\partial J_{r_1}^a}\right]_{X=\bar{X}}
\nonumber\\
&=& 
\sum_r\Big[\frac{\partial^3 W}{\partial X_r \partial J_{r_3}^c \partial J_{r_4}^d} \frac{\partial^2 X_r}{\partial J_{r_1}^a\partial J_{r_2}^b}+\frac{\partial^3 W}{\partial X_r \partial J_{r_2}^b \partial J_{r_4}^d} \frac{\partial^2 X_r}{\partial J_{r_1}^a\partial J_{r_3}^c}+\frac{\partial^3 W}{\partial X_r \partial J_{r_2}^b \partial J_{r_3}^c} \frac{\partial^2 X_r}{\partial J_{r_1}^a\partial J_{r_4}^d}
\nonumber\\
&+&\frac{\partial^3 W}{\partial X_r \partial J_{r_1}^a \partial J_{r_2}^b} \frac{\partial^2 X_r}{\partial J_{r_3}^c\partial J_{r_4}^d}+\frac{\partial^3 W}{\partial X_r \partial J_{r_1}^a \partial J_{r_3}^c} \frac{\partial^2 X_r}{\partial J_{r_2}^b\partial J_{r_4}^d}+\frac{\partial^3 W}{\partial X_r \partial J_{r_1}^a \partial J_{r_4}^d} \frac{\partial^2 X_r}{\partial J_{r_2}^b\partial J_{r_3}^c}\nonumber\\
&+& \sum_{r'} \Big(\frac{\partial^2 W}{\partial X_r \partial X_{r'}} \frac{\partial^2 X_r}{\partial J_{r_1}^a\partial J_{r_3}^c}\frac{\partial^2 X_{r'}}{\partial J_{r_2}^b\partial J_{r_4}^d}+\frac{\partial^2 W}{\partial X_r \partial X_{r'}} \frac{\partial^2 X_r}{\partial J_{r_1}^a\partial J_{r_4}^d}\frac{\partial^2 X_{r'}}{\partial J_{r_2}^b\partial J_{r_3}^c}\nonumber\\
&+&\frac{\partial^2 W}{\partial X_r \partial X_{r'}} \frac{\partial^2 X_r}{\partial J_{r_1}^a\partial J_{r_2}^b}\frac{\partial^2 X_{r'}}{\partial J_{r_3}^c\partial J_{r_4}^d}  \Big)\Big]_{X=\bar{X}}, \nn
\label{eq:4pt}
\end{eqnarray}
where the terms that vanish
at $J^a_r=0$ are dropped.
Defining $\mathbb{L}$ to be a matrix whose element is given by $\mathbb{L}_{rr'}=[({\bf K}+xI)^{-1}_{rr'}]^2+\frac{m^4}{4\lambda} \delta_{rr'}$, 
we write 
\begin{eqnarray}
\frac{\partial^2 X_r}{\partial J_{r_1}^a\partial J_{r_2}^b} \Big|_{\substack{J=0 \\ \vspace{-1.5mm} \\ X=\bar{X}}}&=& \frac{ 2\delta_{ab} }{N m^2 }  \sum_{r'} \mathbb{L}^{-1}_{r r' } ({\bf K}+xI)^{-1}_{r' r_1}({\bf K}+xI)^{-1}_{r_2 r'} \\
\frac{\partial^3 W}{\partial X_r \partial J_{r_1}^a \partial J_{r_2}^b}\Big|_{\substack{J=0 \\ \vspace{-1.5mm} \\ X=\bar{X}}} &=& \frac{\delta_{ab}}{m^2}   ({\bf K}+xI)^{-1}_{r_1,r}({\bf K}+xI)^{-1}_{r,r_2},
\\
\frac{\partial^2 W}{\partial X_r \partial X_{r'}}\Big|_{\substack{J=0 \\ \vspace{-1.5mm} \\ X=\bar{X}}} &=&- \frac{N}{2}\mathbb{L}_{rr'}
\end{eqnarray}
from Eq.~(\ref{eq:Stot}) and Eq.~(\ref{eq:eqforX0}). 
In the momentum space,
$\mathbb{L}$ can be written as
$\mathbb{L}_{rr'}=\int  \frac{d^D q}{(2\pi)^D}e^{i q\cdot(r-r')} \mathbb{L}_q $,
where $\mathbb{L}_{q} = c_1  \int_0^1 du ~[u(1-u)\frac{q^2}{m^2}+x_2e^{-2z}]^{\frac{D}{2}-2} +\frac{m^4}{4\lambda}=c_2 \frac{q^{D-4}}{m^{D-4}}+\frac{m^4}{4\lambda}$ in the large $z$ limit with $c_1=(4\pi)^{-D/2}\Gamma[2-\frac{D}{2}]$ and $c_2=\frac{\Gamma[\frac{4-D}{2}]\Gamma[\frac{D-2}{2}]}{2^{2D-3}\pi^{\frac{D-1}{2}}\Gamma[\frac{D-1}{2}]}$. 
Then, two different terms that appear
Eq.~(\ref{eq:4pt}) are given by
\begin{eqnarray}
-\sum_r \frac{\partial^3 W}{\partial X_r \partial J_{r_3}^c \partial J_{r_4}^d} \frac{\partial^2 X_r}{\partial J_{r_1}^a\partial J_{r_2}^b}\Big|_{\substack{J=0 \\ \vspace{-1.5mm} \\ X=\bar{X}}} &=& - \frac{2\delta_{ab}\delta_{cd}}{N m^4}\sum_{rr'} \mathbb{L}^{-1}_{r r' } ({\bf K}+xI)^{-1}_{r_3,r}({\bf K}+xI)^{-1}_{r,r_4}  ({\bf K}+xI)^{-1}_{r' r_1}({\bf K}+xI)^{-1}_{r_2 r'} \nonumber\\
&=&  \frac{2\delta_{ab}\delta_{cd}}{N m^4} I^z[r_1,r_2,r_3,r_4], \nn
\sum_{r,r'} \frac{\partial^2 W}{\partial X_r \partial X_{r'}} \frac{\partial^2 X_r}{\partial J_{r_1}^a\partial J_{r_2}^c}\frac{\partial^2 X_{r'}}{\partial J_{r_3}^b\partial J_{r_4}^d}
&=&
\frac{2\delta_{ab}\delta_{cd}}{N m^4} I^z[r_1,r_2,r_3,r_4],
\end{eqnarray}
where
\begin{eqnarray}
I^z[r_1,r_2,r_3,r_4] &=&
\begin{tikzpicture}[baseline={([yshift=-4pt]current bounding box.center)}]
\tikzset{decoration={snake,amplitude=.4mm,segment length=1mm, post length=0mm,pre length=0mm}}
\coordinate (v4) at (0pt,12pt);
\coordinate (v3) at (20pt,12pt);
\coordinate (v2) at (0pt,-12pt);
\coordinate (v1) at (20pt,-12pt);
\coordinate (v5) at (10pt, -6pt);
\coordinate (v6) at (10pt,6pt);
\draw[thick](v1)--(v5);
\draw[thick](v2)--(v5);
\draw[thick](v3)--(v6);
\draw[thick](v4)--(v6);
\draw[thick,decorate](v5)--(v6);
\node at (-2pt,15pt) {\scriptsize $r_3$};
\node at (-2pt,-15pt) {\scriptsize $r_1$};
\node at (22pt,15pt) {\scriptsize $r_4$};
\node at (22pt,-15pt) {\scriptsize $r_2$};
\node at (5pt,-4pt) {\scriptsize $r'$};
\node at (5pt,4pt) {\scriptsize $r$};
\end{tikzpicture}=
- \int d^D r d^D r' \frac{1}{|r_1-r'|^{D-2}}\frac{1}{|r_2-r'|^{D-2}}\frac{1}{|r-r'|^{4}}\frac{1}{|r_3-r|^{D-2}}\frac{1}{|r_4-r|^{D-2}}.
\end{eqnarray}
Note that 
$I^z[r_1,r_2,r_3,r_4]$
is a product of four free propagator and a propagator of the O(N) singlet,
and invariant under the interchange of the first two and the last two coordinates, $I^z[r_1,r_2,r_3,r_4]=I^z[r_3,r_4,r_1,r_2]$. 
In the momentum space, it can be expressed as
\begin{eqnarray}
I^z[r_1,r_2,r_3,r_4] &=&-\int \frac{d^D q}{(2\pi)^D}\frac{d^D p}{(2\pi)^D}\frac{d^D Q}{(2\pi)^D}~\mathbb{L}^{-1}_{Q} \frac{e^{i q\cdot (r_3-r_4)}}{\frac{q^2}{m^2}+x_2e^{-2z}}\frac{e^{i Q\cdot (r_4-r_2)}}{\frac{(q-Q)^2}{m^2}+x_2e^{-2z}}      \frac{e^{i p\cdot (r_2-r_1)}}{\frac{p^2}{m^2}+x_2e^{-2z}}\frac{}{\frac{(p-Q)^2}{m^2}+x_2e^{-2z}}\nonumber\\
&=&\frac{m^{8-D}}{4(2\pi)^{D}} \int \frac{d^D Q}{(2\pi)^D} \frac{1}{c_2 \frac{Q^{D-4}}{m^{D-4}}+\frac{m^4}{4\lambda}}\int_0^1 dx \int_0^1 dy \Big(\frac{\sqrt{x(1-x)}\sqrt{y(1-y)}|Q|^2}{r_{12}r_{34}}\Big)^{\frac{D-4}{2}}\nonumber\\
&\times& K_{\frac{D-4}{2}}(\sqrt{x(1-x)}|Q|r_{34})K_{\frac{D-4}{2}}(\sqrt{y(1-y)}|Q|r_{12})e^{i|Q| r_{13} \cos \theta_{13}}e^{ix |Q|r_{34}\cos \theta_{34}}e^{-iy|Q|r_{12} \cos \theta_{12}}.
\end{eqnarray}
where $K_{\nu}(z)$ is the modified Bessel functions of the second kind. In the above expression, 
$r_i-r_j= r_{ij} \hat{r}_{ij}$ where $\hat{r}_{ij}$ is a unit vector connecting points $i$ and $j$. 
$Q \cdot \hat{r}_{ij} = |Q| |r_{ij}|\cos \theta_{ij}$, 
where $\theta_{ij}$ is the angle between $Q$ and $\hat{r}_{ij}$. 
Together with other terms from permutation, 
we can express the connected 4-pt correlation function as
\begin{eqnarray}
G_4^{abcd}[r_1,r_2,r_3,r_4] = 
\frac{2}{Nm^4}
\left[\delta_{ab}\delta_{cd}I^z[r_1,r_2,r_3,r_4]+\delta_{ac}\delta_{bd}I^z[r_1,r_3,r_2,r_4]+\delta_{ad}\delta_{bc}I^z[r_1,r_4,r_2,r_3]\right].
\end{eqnarray}
In general, the connected $4$-point function can be written as 
\begin{eqnarray}
G_4^{abcd}[r_1,r_2,r_3,r_4] = 
f^{abcd}\left(
\frac{ r_{12} r_{34} }{r_{13} r_{24}},
\frac{ r_{12} r_{34} }{r_{14} r_{23}}
\right)
\prod_{i>j} r_{ij}^{-\frac{D-2}{3}},
\end{eqnarray}
where $r_{ij}=|r_i-r_j|$ and 
$f^{abcd}(u,v)$ is a universal function of the cross ratios.
As a concrete example, let us consider a case where the four points lie on a line 
with
$r_{12}=r_{34}=\alpha r$ and $r_{14}=r_{23}=r$.
$\alpha$ is a dimensionless parameter that determines the cross ratios. 
In this case,
the $4$-point function 
in the large $r$ limit becomes
\begin{eqnarray}
G_4^{abcd}[r_1,r_2,r_3,r_4] = 
\frac{2}{N}
\frac{1}{r^{2D-4}} \left(
F_1[\alpha]
+F_2[\alpha]
+F_3[\alpha]
\right),
\end{eqnarray}
where
\begin{eqnarray}
F_1[\alpha] &=&- \frac{1}{4(2\pi)^{D}c_2}\frac{1}{\alpha^{D-4}} \int \frac{d^D Q'}{(2\pi)^D}  \int_0^1 dx \int_0^1 dy \Big(\sqrt{x(1-x)}\sqrt{y(1-y)}\Big)^{\frac{D-4}{2}}\nonumber\\
&\times& K_{\frac{D-4}{2}}(\sqrt{x(1-x)}\alpha |Q'|)K_{\frac{D-4}{2}}(\sqrt{y(1-y)}\alpha|Q'|)
e^{i[(x-y) \alpha +1] |Q'|\cos \theta}, \nn
F_2[\alpha] &=&-\frac{1}{4(2\pi)^{D}c_2} \frac{1}{(1-\alpha^2)^{\frac{D-4}{2}}}\int \frac{d^D Q'}{(2\pi)^D} \int_0^1 dx \int_0^1 dy \Big(\sqrt{x(1-x)}\sqrt{y(1-y)}\Big)^{\frac{D-4}{2}}\nonumber\\
&\times& K_{\frac{D-4}{2}}(\sqrt{x(1-x)}(1+\alpha)|Q'|)K_{\frac{D-4}{2}}(\sqrt{y(1-y)}(1-\alpha)|Q'|)e^{i[\alpha+x(1+\alpha)+y(1-\alpha)] |Q'|\cos \theta}, \nonumber\\
F_3[\alpha] &=&-\frac{1}{4(2\pi)^{D}c_2} \int \frac{d^D Q'}{(2\pi)^D} \int_0^1 dx \int_0^1 dy \Big(\sqrt{x(1-x)}\sqrt{y(1-y)}\Big)^{\frac{D-4}{2}}\nonumber\\
&\times& K_{\frac{D-4}{2}}(\sqrt{x(1-x)}|Q'|)K_{\frac{D-4}{2}}(\sqrt{y(1-y)}|Q'|)e^{i(\alpha+x+y) |Q'|\cos \theta}
\end{eqnarray}
for $D<4$.
Similarly, we can compute other $n$-point functions from the exact effective action.

\section{
Effective action from an alternative RG scheme
\label{app:momentum_space_RG}
}

In this section, 
we derive the fixed point effective action in an alternative RG scheme,
where the free massless theory is used as the reference theory.
The partition function is still written as the overlap between two states as $Z= \langle S_{ref}' | S_1' \rangle$.
The state associated with the new reference theory reads
\begin{eqnarray}
| S'_{ref}\rangle &=& \int \mathcal{D}\phi~e^{-\frac{1}{2} \int d^D k~ G^{-1}_M(k) \phi_k \phi_{-k}}|\phi\rangle,
\end{eqnarray}
where $k$ is momentum
and 
$G^{-1}_M(k) = e^{ \frac{k^2}{M^2} } k^2$
is a regularized kinetic term
that suppresses modes with momenta larger than UV cutoff $M$. 
The deformation includes the bi-linear operators and the on-site quartic interaction,
$S_{1}' = 
\sum_{ij} J_{ij} \phi_{i}\phi_{j}
+
\frac{\lambda}{N}\sum_{i }  ( \phi_{i}^2)^2
$.
The state associated with the deformation at $z=0$ is 
\begin{eqnarray}
| S_1'{}^0 \rangle
&=&\int \mathcal{D}t^0\mathcal{D}p^0~e^{-NS_{UV}[t^0,p^0]}|t^0\rangle,
\end{eqnarray}
where the UV boundary action is $S_{UV}=  \int d^D r d^D r'~(
it^0_{rr'}
+J_{rr'}
)p^0_{rr'}+\lambda \int d^D r~ (p^0_{rr})^2$,
and $| t \rangle$ is defined in \eq{eq:t}.
The RG Hamiltonian that 
satisfies $\hat{H}^\dag |S_{ref}\rangle =0$ 
is given by 
\cite{POLCHINSKI1984exactRG,lee2016horizon}
\begin{eqnarray}
\hat{H} = \int d^D k \left[\frac{\tilde{G}(k)}{2}\hat{\pi}_k \hat{\pi}_{-k}-i \Big(\frac{D+2}{2}\hat{\phi}_k+k\partial_k \hat{\phi}_k\Big)\hat{\pi}_{k}+C\right],
\end{eqnarray}
where $\tilde G(k) = \frac{\partial G_M(k)}{\partial \ln M}=\frac{2}{M^2}e^{-\frac{k^2}{M^2}}$ and $C=-\int d^D k \delta^D (0) \left[\frac{\tilde{G}}{2}G_{M}^{-1}+1\right]$ is a constant. $\hat{\pi}$ is the canonical conjucate of $\hat{\phi}$ satisfying $[\hat{\phi}_k,\hat{\pi}_{k'}]=i (2\pi)^D \delta^D(k-k')$.
The first term in the Hamiltonian  effectively integrates out modes 
with momenta between $M e^{-dz}$ and $M$.
The second term rescales the momentum and field as 
$k=\tilde ke^{-dz}$ and 
$\phi_k =e^{\frac{D+2}{2}dz}\phi_{\tilde{k}}$. 
Since the RG Hamiltonian is diagonal in the momentum space, it is convenient to 
express the basis states
in \eq{eq:t}
in the momentum space as
\begin{eqnarray}
|t\rangle =\int \mathcal{D}\phi e^{i \int d^D k_1 d^D k_2 t_{k_1,k_2}\phi_{k_1}\phi_{k_2}}|\phi\rangle,
\end{eqnarray}
where
$t_{r_1,r_2}=\int d^D k_1 d^D k_2 t_{k_1,k_2} e^{ik_1\cdot r_1+ik_2 \cdot r_2}$
and
$\phi_r =\int \frac{d^D k}{(2\pi)^D}
\phi_k e^{-ik \cdot r}$

An infinitesimal RG transformation applied to the basis state at $z$ can be written as a linear superposition of the basis states at $z+dz$ as 
\begin{eqnarray}
e^{-\hat{H}dz}|t^z\rangle =\int \mathcal{D}t^{z+dz}_{\tilde{k}_1,\tilde{k}_2}\mathcal{D}p^{z+dz}_{\tilde{k}_1,\tilde{k}_2} e^{-N\mathcal{L}^{z+dz}_{bulk}[t^{z+dz}_{\tilde{k}_1,\tilde{k}_2},p^{z+dz}_{\tilde{k}_1,\tilde{k}_2}]dz} |t^{z+dz}\rangle,
\end{eqnarray}
where 
\begin{eqnarray}
\mathcal{L}_{bulk}
&=& i\int d^D \tilde{k}_1 d^D \tilde{k}_2~ (\partial_z t_{\tilde{k}_1,\tilde{k}_2}) p_{\tilde{k}_1,\tilde{k}_2}-i\int d^D  \tilde{k} ~\tilde G( \tilde{k})    t_{ \tilde{k},- \tilde{k}}\nonumber\\
&+&2\int d^D \tilde{k} 
d^D \tilde{k}_1 
d^D \tilde{k}_2 ~ 
~\tilde G(\tilde{k})   
t_{\tilde{k}_1,-\tilde{k}} t_{\tilde{k}_2,\tilde{k}}p_{\tilde{k}_1 ,\tilde{k}_2 }-i\int d^D \tilde{k}_1 d^D \tilde{k}_2~ (2-D)t_{\tilde{k}_1,\tilde{k}_2 } p_{\tilde{k}_1 ,\tilde{k}_2 }\nonumber\\
&+&i\int d^D  \tilde{k}_1 d^D  \tilde{k}_2 ~\tilde{k}_1 (\partial_{\tilde{k}_1}t_{ \tilde{k}_1, \tilde{k}_2 }) p_{\tilde{k}_1,\tilde{k}_2} +i\int d^D  \tilde{k}_1 d^D  \tilde{k}_2 ~\tilde{k}_2 (\partial_{\tilde{k}_2}t_{ \tilde{k}_1, \tilde{k}_2 }) p_{\tilde{k}_1,\tilde{k}_2}.
\end{eqnarray}
Here we use $\tilde k_i$ for momentum to denote the fact that momentum has been already scaled through the RG Hamiltonian.
In the phase space path integration representation,
$| S_1^{'z} \rangle
= e^{-\hat H z} | S_1'{}^0 \rangle$
can be expressed as the path integration over 
$t_{k_1 k_2}^z$,
$p_{k_1 k_2}^z$
as in \eq{eq:S1z},
where the bulk action 
is given by
$S_{bulk}= \int_0^{z^\ast}  dz ~ {\cal L}_{bulk}$.

The integration over $p$ in the bulk gives rise to the constraint for $t$,
\begin{eqnarray}
\partial_z t_{\tilde{k}_1,\tilde{k}_2} -2i\int d^D \tilde{k} ~\tilde G(\tilde{k})    t_{\tilde{k}_1,-\tilde{k}} t_{\tilde{k}_2,\tilde{k}} +  (D-2)t_{\tilde{k}_1,\tilde{k}_2 }  + \tilde{k}_1 (\partial_{\tilde{k}_1}t_{ \tilde{k}_1, \tilde{k}_2 }) + \tilde{k}_2 (\partial_{\tilde{k}_2}t_{ \tilde{k}_1, \tilde{k}_2 }) =0.
\end{eqnarray}
Treating $\{ t_{\tilde k_1 \tilde k_2} \}$  as elements of  
matrix ${\bf t}$,
we can write the solution 
as
\begin{eqnarray}
it_{ \tilde{k}_1,\tilde{k}_2}(z)&=&\Big((i\tilde{\bf t}^0)\left[I - \tilde {\bf D}(z) (i\tilde{\bf t}^0)\right]^{-1}\Big)_{\tilde k_1, \tilde k_2},
\end{eqnarray}
where
$\tilde{t}^0_{\tilde{k}_1,\tilde{k}_2}=e^{(2-D)z}t^0_{\tilde k_1 e^{-z},\tilde k_2  e^{-z}}$, 
and
$\tilde{\bf D}_{\tilde{k},\tilde{k}'}=\tilde{D}_{\tilde{k}}\delta(\tilde{k}+\tilde{k}')$ with $\tilde{D}_{\tilde{k}}=\frac{2}{\tilde{k}^2} \left[\exp(-\frac{\tilde{k}^2e^{-2z} }{M^2})-\exp(- \frac{\tilde{k}^2}{M^2})\right]$. 
We can further integrate over $p^0_{rr'}$ in $S_{UV}$ as $\int \mathcal{D}p_{rr'}^0 e^{-N S_{UV}} 
=\prod_{r\neq r'}\delta(
t^0_{rr'}
-i J_{rr'}
)e^{
-\frac{N }{4\lambda}\int   d^D r~ 
(t^0_{rr} - i  J_{rr})^2}$. 
In the large $N$ limit, the integration over $t^0$ can be replaced with the saddle-point solution.
The saddle-point equation for $t_{rr'}$ is solved if 
we write 
$-i{\bf t}^0
=
{\bf J} +  {\bf X}$,
where ${\bf J}_{k_1,k_2}=J_{k_1}\delta(k_1+k_2)$ is the UV hopping matrix in the momentum space,
and
${\bf X}$ is a matrix whose elements depend only on $k_1+k_2$ as 
${\bf X}_{k_1,k_2}= X_{k_1+k_2}$.
$\tilde{\bf X}_{\tilde k_1,\tilde k_2}
\equiv 
e^{(2-D)z^\ast}{\bf X}_{k_1,k_2}$
satisfies
\begin{eqnarray}
\label{eq:saddle_point_momentum_rescaling}
&&  (2\pi)^D  \frac{N  }{2\lambda } e^{(D-4)z^\ast} 
\tilde{X}_{-\tilde p}
= 
\frac{N}{2 }  \int d^D \tilde k_1 ~\tilde{ D}_{\tilde k_1} \left[ I+
\tilde{\bf D} 
(\tilde{\bf X}+\tilde{\bf J})\right]^{-1}_{\tilde{p}+\tilde k_1,\tilde k_1}  \\
&+& \int d^D \tilde k_1 \Big\{ \int d^D   \tilde{k} ~ \phi_{ \tilde{k} }    
\left[I +  (\tilde{\bf X}+\tilde{\bf J})\tilde{\bf D}\right]^{-1}_{\tilde k,\tilde k_1} \Big\}\Big\{ \int d^D   \tilde{k}'
\left[I + \tilde{\bf D} (\tilde{\bf X}+\tilde{\bf J})\right]^{-1}_{\tilde p-\tilde k_1,\tilde k'}  \phi_{ \tilde{k}'}\Big\},
\nonumber
\end{eqnarray}
where $\tilde{\bf J}_{\tilde k_1,\tilde k_2}
\equiv 
e^{(2-D)z^\ast}{\bf J}_{k_1,k_2}$.
The effective action at scale $z^\ast$ is written as
\begin{eqnarray}
\label{eq:partition_function_momentum_rescaling}
S_{tot}
&=&
-(2 \pi)^D \frac{N }{4 \lambda}e^{(D-4)z^\ast}
\int   d^D \tilde{p}~ \tilde{ X}_{\tilde p} \tilde{ X}_{-\tilde p} 
+\frac{N}{2}\int d^D \tilde{k}~ 
\log \left[ I+\tilde{\bf D}(
\tilde{\bf X}+\tilde{\bf J})\right]_{\tilde {k},\tilde {k}}  
\\ && 
+\frac{1}{2} \int d^D \tilde{k}~ G^{-1}_{M}(\tilde{k}) \phi_{\tilde{k}} \phi_{-\tilde{k}}   
+ \int d^D \tilde{k}_1 d^D   \tilde{k}_2~ \Big((\tilde{\bf X}+\tilde{\bf J})\left[I + \tilde{\bf D} (\tilde{\bf X}+\tilde{\bf J})\right]^{-1}\Big)_{\tilde{k}_1,\tilde{k}_2}\phi_{ \tilde{k}_1 }\phi_{ \tilde{k}_2}. \nonumber
\end{eqnarray}
If we write $\phi$-independent part of $\tilde {\bf X}$ as $\tilde {\bf x}$, it satisfies 
\begin{eqnarray}
&& \int^{\Lambda e^{z^\ast}} d^D \tilde k_1 \tilde{ D}_{\tilde k_1} \left[ I+
\tilde{\bf D} 
(\tilde{\bf x}+\tilde{\bf J})\right]^{-1}_{\tilde p+\tilde k_1,\tilde k_1} 
=
\frac{(2\pi)^D}{  \lambda } 
e^{(D-4)z^\ast}   \tilde{x}_{-\tilde p},
\end{eqnarray}
where
$(\tilde{\bf x})_{\tilde k_1,\tilde k_2}=\tilde{ x}_{\tilde k_1+\tilde k_2}$
and
$\Lambda$ is the UV cutoff at $z=0$. 
In the presence of the translational symmetry,
$\tilde{ x}_{-\tilde p}
\propto
\delta(
\tilde p)
$.
$\tilde {\bf x}$ can be expanded as
$\tilde {\bf x} = 
\tilde {\bf x}_0 e^{2z^\ast}
+ \tilde {\bf x}_2 
+ O(e^{-2z^\ast})
$.
At the critical point,
the exponentially growing part of 
$\tilde {\bf x}$
cancels with that 
of 
$\tilde {\bf J}$,
and
$\tilde{\bf x}+\tilde{\bf J} \sim \mathcal{O}(1)$.
If we define
$\tilde{\bf X}'
\equiv \tilde {\bf X} - 
\tilde {\bf x}_0 e^{2z^\ast}
$,
$\tilde {\bf X}'$ satisfies
\begin{eqnarray}
\label{eq:saddle_point_X_prime}
&&  \frac{N}{2}  \int d^D\tilde{k}_1  ~\tilde{ D}_{\tilde{k}_1} \left[ \tilde{\bf T}+
\tilde{\bf D} 
\tilde{\bf X}'\right]^{-1}_{\tilde{p}+\tilde{k}_1,\tilde{k}_1} 
-\frac{N}{2}  \int d^D\tilde{k}_1  ~
\tilde{ D}_{\tilde{k}_1}  
\left[
\tilde{\bf T} +
\tilde {\bf D} \tilde x_2
\right]^{-1}_{\tilde{p}+\tilde{k}_1,\tilde{k}_1}  \\
&+& \int d^D\tilde{k}_1 \Big\{\int d^D  \tilde{k} ~    \phi_{ \tilde{k} } \left[\tilde{\bf T}^T +   \tilde{\bf X}'\tilde{\bf D}\right]^{-1}_{\tilde{k},\tilde{k}_1} \Big\}\Big\{\int d^D  \tilde{k}'\left[\tilde{\bf T} + \tilde{\bf D}  \tilde{\bf X}'\right]^{-1}_{\tilde{p}-\tilde{k}_1,\tilde{k}'}  \phi_{ \tilde{k}'} \Big\}\nonumber\\
&=&(2\pi)^D\frac{N }{2\lambda } e^{(D-4)z^\ast}    (\tilde{X}'_{-\tilde{p}}
- \tilde {\bf x}_2), \nonumber
\end{eqnarray}
where
$\tilde{\bf T}=I+ \tilde{\bf D}  (\tilde{\bf x}_0 e^{2 z^\ast} +\tilde{\bf J})$. 
When $D>4$, 
$\tilde{\bf X}'=  {\bf x}_2$,
and
the fixed point action is Gaussian.
For $D<4$, $\tilde{\bf X}'$ becomes $\phi$-dependent,
and Eq.~(\ref{eq:partition_function_momentum_rescaling}) becomes non-Gaussian.

If we deform the UV theory 
by adding  a uniform mass 
$ \delta J_{k_1 k_2} = \epsilon' \delta(k_1+k_2)$,
which translates to
$ \delta \tilde J_{\tilde k_1 \tilde k_2} = \epsilon' e^{2z} \delta(\tilde k_1+\tilde k_2)$,
the effective action changes by
\begin{eqnarray}
\Delta_\epsilon' S_{tot} 
&\propto & 
\epsilon'
\left[
e^{Dz} \tilde x_0
+
e^{(D-2)z} \tilde X'_0
\right].
\end{eqnarray}
This implies
that 
$\tilde {\cal X}'_{\tilde r}
\equiv 
\int d^D \tilde k
\tilde X'_{\tilde k} e^{i \tilde k \tilde r}
$
is a local operator
with with scaling dimension $\Delta_X=2$.
If a non-local hopping term is turned on between $r$ and $r'$ at UV with strength $\epsilon$,
$\delta J_{k_1 k_2} = \epsilon e^{-i k_1 r - i k_2 r'}$
($\delta \tilde J_{\tilde k_1 \tilde k_2} = 
\epsilon 
e^{(2-D)z}
e^{-i \tilde k_1 \tilde r - i \tilde k_2 \tilde r'}$),
the change of the effective action is given by
\begin{eqnarray}
\Delta_\epsilon S_{tot} &\propto& 
\epsilon e^{(2-D)z}
\int d^D  \tilde k  
d^D  \tilde k'
~
e^{-i 
(\tilde k\cdot \tilde r
+
\tilde k' \cdot \tilde r'
)
}  
\phi^S_{\tilde k}
\phi^S_{\tilde k'}
\label{eq:action_diff_momentum_phi}
\end{eqnarray}
in the   $|\tilde r-\tilde r'| \rightarrow \infty$ limit,
where
$
\phi^S_{\tilde k}=\int d^D \tilde{k}'  \left[I + \tilde{\bf D} (\tilde{\bf X}+\tilde{\bf J})\right]^{-1}_{\tilde k',\tilde k} \phi_{\tilde{k}'}$.
$\Phi^S_{\tilde r}
\equiv
\int d^D  \tilde k  
~
e^{-i  \tilde k\cdot \tilde r}
\phi^S_{\tilde k}$
corresponds to the local scaling operator in the fundamental representation of $O(N)$
with scaling dimension
$\Delta_\phi = \frac{D-2}{2}$.

\end{document}